\documentclass[aps,pra,reprint,superscriptaddress,10pt, longbibliography, nofootinbib]{revtex4-1}
\usepackage{amsmath}
\usepackage{hyperref}
\usepackage{graphicx}
\usepackage[dvipsnames]{xcolor}
\usepackage{pifont}

\DeclareMathAlphabet\mathbfcal{OMS}{cmsy}{b}{n}

\begin{document}

\title{Micromotion minimization using Ramsey interferometry}

\author{Gerard Higgins}
\email[]{higgins@chalmers.se. Present address: Institute for Quantum Optics and Quantum Information (IQOQI), Austrian Academy of Sciences, Vienna, Austria}
\affiliation{Department of Physics, Stockholm University, Stockholm, Sweden}
\affiliation{Department of Microtechnology and Nanoscience, Chalmers University of Technology, Gothenburg, Sweden}
\author{Shalina Salim}
\affiliation{Department of Physics, Stockholm University, Stockholm, Sweden}
\author{Chi Zhang}
\email[]{Present address: Centre for Cold Matter, Imperial College London, London, UK}
\affiliation{Department of Physics, Stockholm University, Stockholm, Sweden}
\author{Harry Parke}
\affiliation{Department of Physics, Stockholm University, Stockholm, Sweden}
\author{Fabian Pokorny}
\email[]{Present address: Department of Physics, University of Oxford, Oxford, UK}
\affiliation{Department of Physics, Stockholm University, Stockholm, Sweden}
\author{Markus Hennrich}
\affiliation{Department of Physics, Stockholm University, Stockholm, Sweden}

\date{\today}

\let\vec\mathbf

\begin{abstract}
We minimize the stray electric field in a linear Paul trap quickly and accurately, by applying interferometry pulse sequences to a trapped ion optical qubit.
The interferometry sequences are sensitive to the change of ion equilibrium position when the trap stiffness is changed, and we use this to determine the stray electric field.
The simplest pulse sequence is a two-pulse Ramsey sequence, and longer sequences with multiple pulses offer a higher precision.
The methods allow the stray field strength to be minimized beyond state-of-the-art levels, with only modest experimental requirements.
Using a sequence of nine pulses we reduce the 2D stray field strength to $(10.5\pm0.8)\,\mathrm{mV\,m^{-1}}$ in 11\,s measurement time.
The pulse sequences are easy to implement and automate, and they are robust against laser detuning and pulse area errors.

We use interferometry sequences with different lengths and precisions to measure the stray field with an uncertainty below the standard quantum limit.
This marks a real-world case in which quantum metrology offers a significant enhancement.
Also, we minimize micromotion in 2D using a single probe laser, by using an interferometry method together with the resolved sideband method; this is useful for experiments with restricted optical access.

Furthermore, a technique presented in this work is related to quantum protocols for synchronising clocks; we demonstrate these protocols here.
\end{abstract}

\maketitle

\section{Introduction}

In a Paul trap ions are confined using an oscillating electric quadrupole field.
Ideally the equilibrium position of a single trapped ion will coincide with the null of the oscillating quadrupole field.
Stray electric fields as well as trap fabrication imperfections introduce a quasi-static dipole electric field $\vec{E}$ at the null of the oscillating quadrupole field, which displaces the ion equilibrium position from the oscillating field null.
This results in an oscillating dipole field at the ion equilibrium position, which drives oscillatory ion motion, called excess micromotion \cite{Berkeland1998}.

The oscillating dipole field causes a Stark shift and the excess micromotion causes a Doppler shift, both effects impact precision spectroscopy \cite{Keller2015}, and the Stark shifts are particularly troublesome in experiments using highly-polarizable Rydberg ions \cite{Higgins2019, Feldker2015}.
Furthermore, the energy stored in excess micromotion is an obstacle to studies of quantum interactions in hybrid systems of neutral atoms and trapped ions \cite{Grier2009,Schmid2010,Zipkes2010,Feldker2020}.
The Stark shift and the excess micromotion can be diminished by applying a static electric dipole field to counter the unwanted quasi-static dipole field $\vec{E}$.
This opposing electric field is usually produced by applying voltages to dedicated compensation electrodes.

Although a host of techniques have been developed to determine appropriate compensation electrode voltages \cite{Berkeland1998, Keller2015, Feldker2020, Barrett2003, Allcock2010, Chuah2013, Schneider2005, Gloger2015, Brown2007,   Ibaraki2011, Narayanan2011, Tanaka2012,  Harter2013, Mohammadi2019, Yu1994, Higgins2019, Cerchiari2020, Zhukas2020}, there is a demand to improve upon the existing techniques.
For instance, the world's most precise clock is currently a trapped ion optical clock \cite{Brewer2019}, and the largest contribution to its systematic uncertainty arises from excess micromotion.

Some of the most popular methods for minimising excess micromotion rely on the impact of micromotion on an ion's absorption or emission spectra, through the Doppler effect \cite{Berkeland1998, Keller2015, Barrett2003, Allcock2010, Chuah2013}.
For instance, micromotion introduces spectral sidebands which are separated from carrier transitions by the frequency of the trap's oscillating quadrupole field \cite{Berkeland1998, Keller2015}.
It also modulates the ion's scattering rate at the frequency of the trap's oscillating quadrupole field \cite{Berkeland1998, Keller2015}.

\begin{figure}[ht]
\centering
\includegraphics[width=\columnwidth]{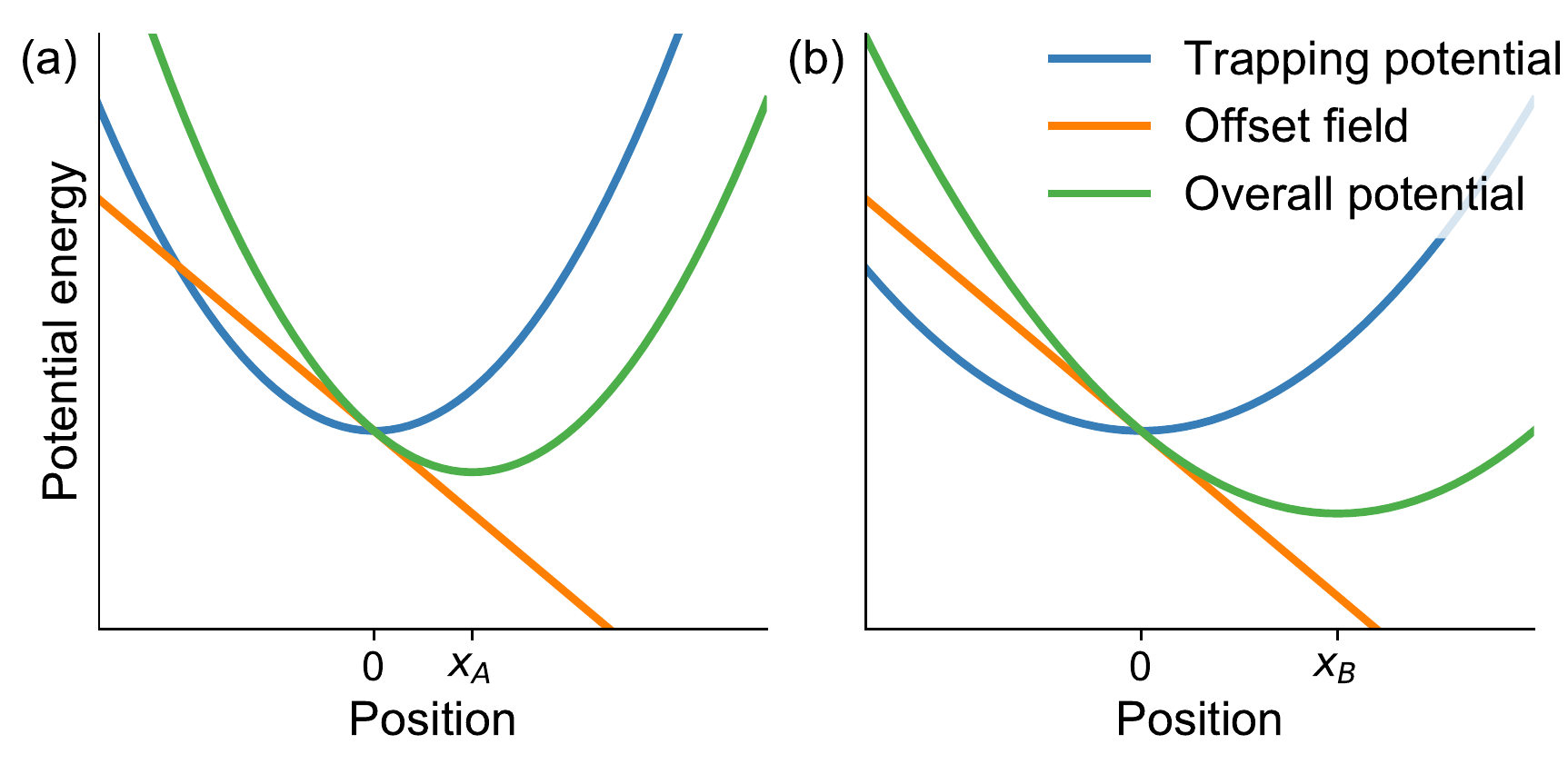}
\caption{At the minimum (position 0) of a Paul trap's effective RF potential (blue lines) there is no oscillating dipolar electric field.
A static dipolar offset field $\vec{E}$ (corresponding to the orange potential) displaces the trapping pseudopotential, causing an ion trapped at the displaced minimum [$x_A$ in (a), $x_B$ in (b)] to experience an oscillating dipolar electric field which drives excess micromotion.
The offset field $\vec{E}$ is the same in both plots.
The displacement from 0 is larger in (b) than in (a) because the trap stiffness is weaker in (b).
By changing the trap stiffness and measuring the change of a trapped ion's position information is gained about $\vec{E}$ [see Eq.~(\ref{eq_del_u})].
}
\label{fig_pseudopotential}
\end{figure}

Other techniques rely on measuring the change of a trapped ion's equilibrium position when the trap stiffness is changed \cite{Berkeland1998, Gloger2015, Schneider2005, Brown2007, Feldker2020, Saito2021}; the methods we present here also work in this fashion.
These techniques are explained as follows:
The unwanted quasi-static dipole field $\vec{E}$ at the position of the trap's oscillating field null displaces the equilibrium position of a trapped ion from the null by $\vec{r}$, where \cite{Berkeland1998}
\begin{equation} \label{eq_r}
r_i = \frac{q E_i}{m {\omega_i}^2}
\end{equation}
and $q$ is the ion charge, $m$ is the ion mass, the three spatial directions indexed by $i$ are defined by the ion's secular motion, and $\omega_i$ is the trap stiffness (the frequency of the trapping pseudopotential) in the $i$ direction.
When the trap stiffness is changed $\omega_{Ai} \rightarrow \omega_{Bi}$ the ion equilibrium position is displaced by $\vec{r}_{AB}$, which has the components
\begin{equation} \label{eq_del_u}
r_{ABi} = \frac{q E_i}{m} \left( \frac{1}{{\omega_{Bi}}^2} - \frac{1}{{\omega_{Ai}}^2} \right)
\end{equation}
This is represented in Fig.~\ref{fig_pseudopotential}.
The trap stiffness is usually changed by altering the amplitude of the trap's oscillating electric quadrupole field, though it can also be changed by altering the amplitude of the trap's static quadrupole field \cite{Gloger2015, Schneider2005}.

By measuring effects sensitive to $\vec{r}_{AB}$ ion trappers gain information about $\vec{E}$.
The displacement $\vec{r}_{AB}$ is commonly monitored by imaging a trapped ion \cite{Berkeland1998, Gloger2015, Schneider2005, Feldker2020, Saito2021}.
It can also be detected by measuring the strength with which transitions are driven when there is an optical field gradient \cite{Brown2007} or a magnetic field gradient \cite{Feldker2020}.
These methods are limited by the imaging resolution, by optical diffraction limits and laser powers, and by achievable magnetic field gradients respectively.

In this work we use interferometry to measure $\vec{r}_{AB}$ with a resolution much less than an optical wavelength.
This allows us to reduce $|\vec{E}|$ beyond state-of-the-art levels in a short time, and thereby diminish excess micromotion.
We apply different Ramsey-interferometry pulse sequences to a single trapped ion to probe $\vec{r}_{AB}$.
Using a sequence of two $\pi/2$ pulses resonant to an optical transition we determine the projection of $\vec{r}_{AB}$ along one direction with resolution $\approx \tfrac{\lambda}{2\pi \sqrt{N}}$, where $\lambda$ is the wavelength of the laser field and $N$ is the number of experimental cycles.
We improve on this resolution using sequences of $M+1$ coherent pulses, which offer a $M$-fold precision enhancement.
The pulse sequences are described in Section~\ref{sec_seqs}.

In Section~\ref{sec_fast_accurate} we demonstrate fast and accurate minimization of $\vec{E}$, and discuss the impact that changing the RF power supplied to the trap has on the trap temperature. 

In Section~\ref{sec_efficient_phase_estimation} we show that by measuring using pulse sequences of different lengths $\vec{r}_{AB}$ and $\vec{E}$ can be probed with an uncertainty below the standard quantum limit.
The pulse sequences can be designed so that the results are robust against pulse area errors and laser detuning; we demonstrate this in Section~\ref{sec_dynamical_decoupling}.

In Section~\ref{sec_2D_and_3D_main} we apply the methods to minimize micromotion in 2D and 3D.
We also demonstrate 2D micromotion minimization using just a single laser beam, by using the interferometry method together with the commonly-used resolved sideband technique \cite{Berkeland1998}.

As well as enabling micromotion minimization, one of the pulse sequences presented here demonstrates clock-synchronization protocols which involve exchange of a ticking qubit \cite{Chuang2000, deBurgh2005}.
This is described in Section~\ref{sec_ticking_qubit}.

\section{Pulse sequences} \label{sec_seqs}
In this section we present methods to minimize $|\vec{E}|$ using interferometry sequences, but first we introduce some key concepts:
The action of a sequence of laser pulses on a transition $|g\rangle \leftrightarrow |e\rangle$ between two states of an ion can be described by a sequence of rotations on the Bloch sphere spanned by $|g\rangle$ and $|e\rangle$. 
When the laser field driving the pulses is resonant to the $|g\rangle \leftrightarrow |e\rangle$ transition, the rotation axes lie on the Bloch sphere's equator.
The phase of the laser field during each pulse, within the ion's rotating frame, determines the azimuthal angle of each rotation axis.

Within the ion's rotating frame, the phase of the laser field is fixed in time (unless a controlled phase shift is introduced), and it varies in space according to
\begin{equation} \label{eq_Phi_alpha_A}
\Phi_{\alpha A} = \vec{k}_\alpha \cdot \vec{r}_A + \Phi_{\alpha 0}
\end{equation}
where $\vec{k}_\alpha$ is the wavevector of the laser field, $\Phi_{\alpha 0}$ is a constant phase offset, and Greek letters are used to index different laser beams along different directions while Roman letters are used to index different trap stiffness settings and the corresponding ion positions.
The laser phase experienced by the ion depends on the ion position.
This means the rotation axis of a laser pulse and the impact the pulse has on the ion's state also depend on the ion's position.
By applying a sequence of pulses and measuring the ion's state we can probe the change of ion position $\vec{r}_{AB}$ when the trap stiffness is changed from setting $A \rightarrow B$.

We use Ramsey pulse sequences, comprising two $\pi/2$ pulses, as well as longer sequences with several $\pi$ pulses between two $\pi/2$ pulses.
In general the sequences comprise $M+1$ pulses and have pulse areas $M \pi$, where $M$ is an integer and $M \geq 1$.

During the pulse sequences the phase of the laser field at the ion position is changed between pulses.
This is accomplished by changing the phase of the laser beam which drives the pulse, or by using a different laser beam from a different direction, or by moving the ion from one position to another.
We write the laser phase experienced by the ion during the $j^\mathrm{th}$ pulse as $\phi_j+\theta_j$, where $\phi_j$ depends on both the ion position and the laser beam used to drive the pulses according to Eq.~(\ref{eq_Phi_alpha_A}), while the controlled shift $\theta_j$ results from adding a phase shift to the laser field, using, for example, an acousto-optical modulator.
$\{\phi_j\}$ are general phases, later we will substitute in specific phases using Eq.~(\ref{eq_Phi_alpha_A}).
If the ion is initially in state $|g\rangle$, after applying the pulse sequence the probability of measuring the ion in state $|e\rangle$ is
\begin{equation} \label{eq_pe_cos_theta_T_phi_T}
p = \tfrac{1}{2} \left[ 1 + \cos{\left( \phi_\mathrm{T} + \theta_\mathrm{T} \right)}\right]
\end{equation}
where
\begin{align} \label{eq_phi_T}
\phi_\mathrm{T}&=\phi_1 + 2\sum_{j=2}^{M} (-1)^{j-1}  \phi_j + (-1)^M \phi_{M+1} \\
\theta_\mathrm{T}&=\theta_1 + 2\sum_{j=2}^{M} (-1)^{j-1}  \theta_j + (-1)^M \theta_{M+1} + \xi_M \label{eq_theta_T}
\end{align}
and where $\xi_M=\pi$ ($0$) if $M$ is even (odd).
The phase $\phi_\mathrm{T}$ reveals information about the ion position, or change of position.
By repeatedly applying the sequence and measuring the state of the ion, the probability $p$ can be estimated, from which $\phi_\mathrm{T}$ can be estimated (the controlled phase shift $\theta_\mathrm{T}$ is known).
An estimate of $\phi_\mathrm{T}$ using a single $p$ estimate and Eq.~(\ref{eq_pe_cos_theta_T_phi_T}) is sensitive to pulse area errors and decoherence.
More robust estimates of $\phi_\mathrm{T}$ use two measurements of $p$ using two different $\theta_\mathrm{T}$ values.
One can use \cite{Chwalla2009}
\begin{equation}
\phi_\mathrm{T} = \arcsin{\frac{p(\theta_\mathrm{T}=-\tfrac{\pi}{2})-p(\theta_\mathrm{T}=\tfrac{\pi}{2})}{\mathcal{C}\left[p(\theta_\mathrm{T}=-\tfrac{\pi}{2})+p(\theta_\mathrm{T}=\tfrac{\pi}{2})\right]}} \label{eq_phiT_2}
\end{equation}
where $\mathcal{C}$ accounts for reduction of the contrast of the oscillation in Eq.~(\ref{eq_pe_cos_theta_T_phi_T}), or one can use the two-argument arctangent function \cite{Kimmel2015}
\begin{equation}
\phi_\mathrm{T} = \mathrm{arctan2}\left[ p(\theta_\mathrm{T}=-\tfrac{\pi}{2})-\tfrac{1}{2}, p(\theta_\mathrm{T}=0)-\tfrac{1}{2} \right] \label{eq_phiT_3}
\end{equation}
Eq.~(\ref{eq_phiT_2}) performs well when $\phi_\mathrm{T}\approx 0$, and returns an estimate within a range of $\pi$, while Eq.~(\ref{eq_phiT_3}) returns an estimate of $\phi_\mathrm{T}$ within a range of $2\pi$.
When $N$ experimental runs are conducted, $\tfrac{N}{2}$ using each value of $\theta_\mathrm{T}$, the statistical uncertainties of the $\phi_\mathrm{T}$ estimates are $\approx \tfrac{1}{\sqrt{N}}$; the statistical uncertainties depend on the magnitude of $\phi_\mathrm{T}$, as shown in Appendix~\ref{appendix_a}.

Pulse area errors and detuning of the laser field from resonance introduce systematic errors to estimates of $\phi_\mathrm{T}$.
Systematic errors can be reduced by appropriately choosing the control phases $\{\theta_j\}$, as shown in Section~\ref{sec_dynamical_decoupling}.

The pulse sequences presented here build on the sequence presented in ref.~\cite{Kimmel2015}.
In Method~A the coherent pulses are driven using a single laser beam and the trap stiffness is changed between pulses.
In Method~B two laser beams are used and the trap stiffness is not changed between coherent pulses.
In Appendix~\ref{appendix_method_c} we describe Method~C, which involves multiple laser beams with trap stiffness changes between the pulses.

\subsection*{Method~A: Sequence using a single laser beam}
\begin{figure}[ht]
\centering
\includegraphics[width=\columnwidth]{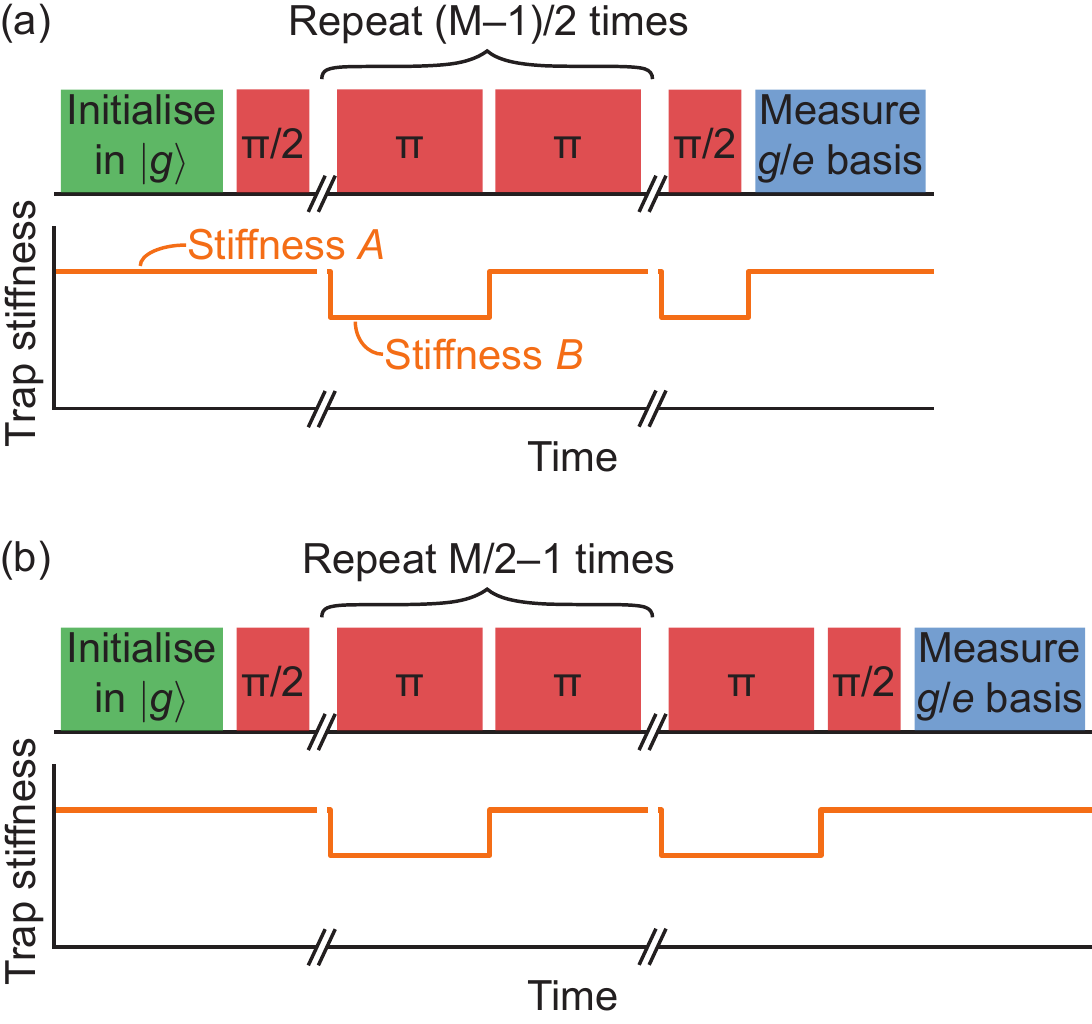}
\caption{Method~A involves a sequence in which the trap stiffness is changed between each coherent laser pulse.
If there is an unwanted field $\vec{E}$, changing the trap stiffness causes the ion to change position and experience a different laser phase.
The probability of measuring the ion in $|e\rangle$ depends on the laser phases during the pulses, and thus on $\vec{E}$.
The areas of the coherent pulses (red) are indicated.
The sequences in (a) and (b) are used when $M$ is odd and even respectively.
The shortest sequence is a Ramsey sequence with $M=1$.
}
\label{fig_method_A}
\end{figure}

In the first method the laser pulses are driven by a single laser beam and the trap stiffness is alternated between stiffness $A$ and stiffness $B$ between laser pulses.
The sequence is presented in Fig.~\ref{fig_method_A}.
The trap stiffness changes cause the ion position to alternate between two positions, $\vec{r}_A$ and $\vec{r}_B$, and the position-dependent phase $\phi_j$ alternates between two values $\Phi_{\alpha A}$ and $\Phi_{\alpha B}$.
Using Eq.~(\ref{eq_Phi_alpha_A}), the difference between the phase values is
\begin{equation} \label{eq_Phi_alpha_A_Phi_alpha_B}
\Phi_{\alpha A} - \Phi_{\alpha B} = \vec{k}_\alpha \cdot \left( \vec{r}_A - \vec{r}_B \right)
\end{equation}
and from Eqs.~(\ref{eq_phi_T}) and (\ref{eq_del_u})
\begin{align}  \label{phi_T_phi_alphaA_phi_alphaB}
\phi_\mathrm{T} &= M \left( \Phi_{\alpha A} - \Phi_{\alpha B} \right) \\ \label{eq_phi_T_k_r_AB}
&= M \vec{k}_\alpha \cdot \left( \vec{r}_A - \vec{r}_B \right) \\
&= M \sum_i \frac{q k_{\alpha i} E_i}{m} \left( \frac{1}{{\omega_{Ai}}^2} - \frac{1}{{\omega_{Bi}}^2} \right) \label{eq_phiT_E}
\end{align}
From Eq.~(\ref{eq_phi_T_k_r_AB}) we see $\phi_\mathrm{T}$ reveals the change in equilibrium position along the direction of $\vec{k}_\alpha$, and from Eq.~(\ref{eq_phiT_E}) we see $\phi_\mathrm{T}$ is sensitive to $\vec{E}$ along the direction $\vec{d}$, which has the components
\begin{equation} \label{eq_d}
d_i = k_{\alpha i} \left( \frac{1}{{\omega_{Ai}}^2} - \frac{1}{{\omega_{Bi}}^2} \right)
\end{equation}
Thus, by probing and minimizing $\phi_\mathrm{T}$, $\vec{E}$ can be minimized.

For convenience we define $\phi_\mathrm{PD} \equiv \Phi_{\alpha A} - \Phi_{\alpha B}$; the phase difference $\phi_\mathrm{PD}$ depends on the path length difference from the laser source to $\vec{r}_A$ and from the laser source to $\vec{r}_B$.
From Eqs.~(\ref{eq_pe_cos_theta_T_phi_T}), (\ref{phi_T_phi_alphaA_phi_alphaB}) and (\ref{eq_phiT_E})
\begin{align} \label{eq_pe_M_phiPD}
p & = \tfrac{1}{2}\left[1+\cos{\left(M\phi_\mathrm{PD}+\theta_\mathrm{T}\right)}\right] \\
\label{eq_pe_M_E} & = \tfrac{1}{2} \left\{1+ \cos{\left[ M \sum_i\frac{q k_{\alpha i} E_i}{m} \left( \frac{1}{{\omega_{Ai}}^2} - \frac{1}{{\omega_{Bi}}^2} \right) + \theta_\mathrm{T} \right]} \right\}
\end{align}
With increasing $M$ the precision of a $\phi_\mathrm{PD}$ estimate is improved, at the expense of reducing the range within which $\phi_\mathrm{PD}$ can be determined.
$\phi_{\mathrm{PD}}$ can be efficiently determined with a Heisenberg scaling by conducting measurements using different values of $M$; this is discussed further in Section~\ref{sec_efficient_phase_estimation}.

We experimentally demonstrate the workings of this method using a single $\mathrm{^{88}Sr^+}$ ion confined in a linear Paul trap.
A 674\,nm laser field couples a Zeeman sublevel of the $5^2S_{1/2}$ ground state $|g\rangle$ with a Zeeman sublevel of the metastable $4^2D_{5/2}$ state $|e\rangle$.
To initialise the ion in $|g\rangle$ we employ Doppler cooling as well as optical pumping on a transition between $5^2S_{1/2}$ and $4^2D_{5/2}$ sublevels.
In some experiments we also employ sideband cooling.
State detection involves probing the ion with 422\,nm laser light near-resonant to the $5^2S_{1/2} \leftrightarrow 5^2P_{1/2}$ transition.
The trap stiffness is changed between the laser pulses by changing the amplitude of the RF signal applied to the trap electrodes and thus changing the amplitude of the trap's oscillating quadrupole field.
The electronics are described in detail in Appendix~\ref{sec_electronics}.

\begin{figure}[ht]
\centering
\includegraphics[width=\columnwidth]{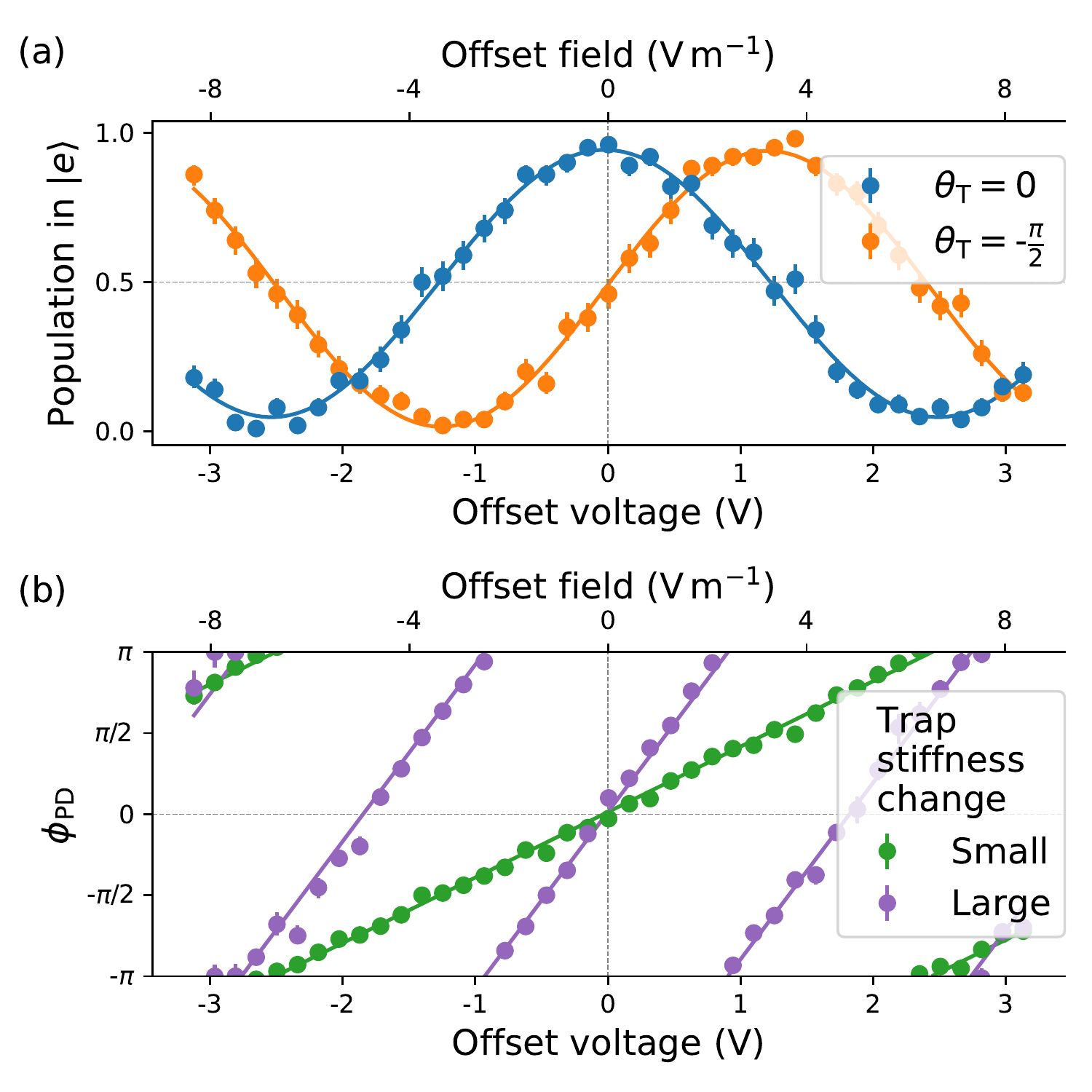}
\caption{Micromotion minimization using Method~A.
(a)~The population measured in $|e\rangle$ depends sinusoidally on the offset voltage applied to a micromotion compensation electrode and the offset field strength.
(b)~The phase difference $\phi_\mathrm{PD}$ depends linearly on the offset voltage, and is zero when micromotion is minimized.
The green data was calculated from the datasets in (a) using Eq.~(\ref{eq_phiT_3}).
$\phi_\mathrm{PD}$ responds more strongly to the offset field when the trap stiffness is changed by a larger amount.
The solid lines in (a) and (b) are respectively sinusoidal and linear fits to the data.
Error bars represent quantum projection noise (1$\sigma$ confidence interval).
The error bars are often smaller than the marker size.
}
\label{fig_seq1}
\end{figure}

A component of $\vec{E}$ is varied by changing the voltage applied to a compensation electrode, and the effect on $p$ is measured in a two-pulse Ramsey sequence ($M=1$).
The results are shown in Fig.~\ref{fig_seq1}(a).
As expected from Eq.~(\ref{eq_pe_M_E}) $p$ shows a sinusoidal dependence on the changes made to $\vec{E}$.

Fig.~\ref{fig_seq1}(a) shows $p$ values when two different values of $\theta_\mathrm{T}$ were used.
From this data and using Eq.~(\ref{eq_phiT_3}) $\phi_\mathrm{PD}$ was calculated; the results are shown in Fig.~\ref{fig_seq1}(b).
The figure shows that $\phi_{\mathrm{PD}}$ has a linear dependence on a component of $\vec{E}$, and that $\phi_{\mathrm{PD}}=0$ when the compensation electrode offset voltage is zero.
The point where the offset voltage is zero was independently determined using the resolved sideband technique \cite{Berkeland1998}.
Throughout this work compensation electrode offset voltages are shown relative to the optimal values as determined using the resolved sideband method.

The figure also shows the linear dependence of $\phi_{\mathrm{PD}}$ on a component of $\vec{E}$ is stronger when the change of the trap stiffness is larger, as expected from the $({\omega_{Ai}}^{-2}-{\omega_{Bi}}^{-2})$ term in Eq.~(\ref{eq_phiT_E}).
The measurements involved reducing the radial secular frequencies from ${\sim2}\pi\times 1.5\,\mathrm{MHz}$ to ${\sim2}\pi\times 600\,\mathrm{kHz}$ for the green dataset and to ${\sim2}\pi\times 400\,\mathrm{kHz}$ for the purple dataset. The axial secular frequency was fixed ${\sim2}\pi\times1.0\,\mathrm{MHz}$.
Because Eq.~(\ref{eq_pe_M_E}) is cyclic it is possible to achieve $\phi_{\mathrm{PD}}=0$ when $|\vec{E}|$ is not minimized, as seen for the purple dataset near $\pm 2\,\mathrm{V}$.
To check that $|\vec{E}|$ is truly minimized, one can check that $\phi_{\mathrm{PD}}$ remains zero when different trap stiffness changes are used.

\begin{figure}[ht]
\centering
\includegraphics[width=\columnwidth]{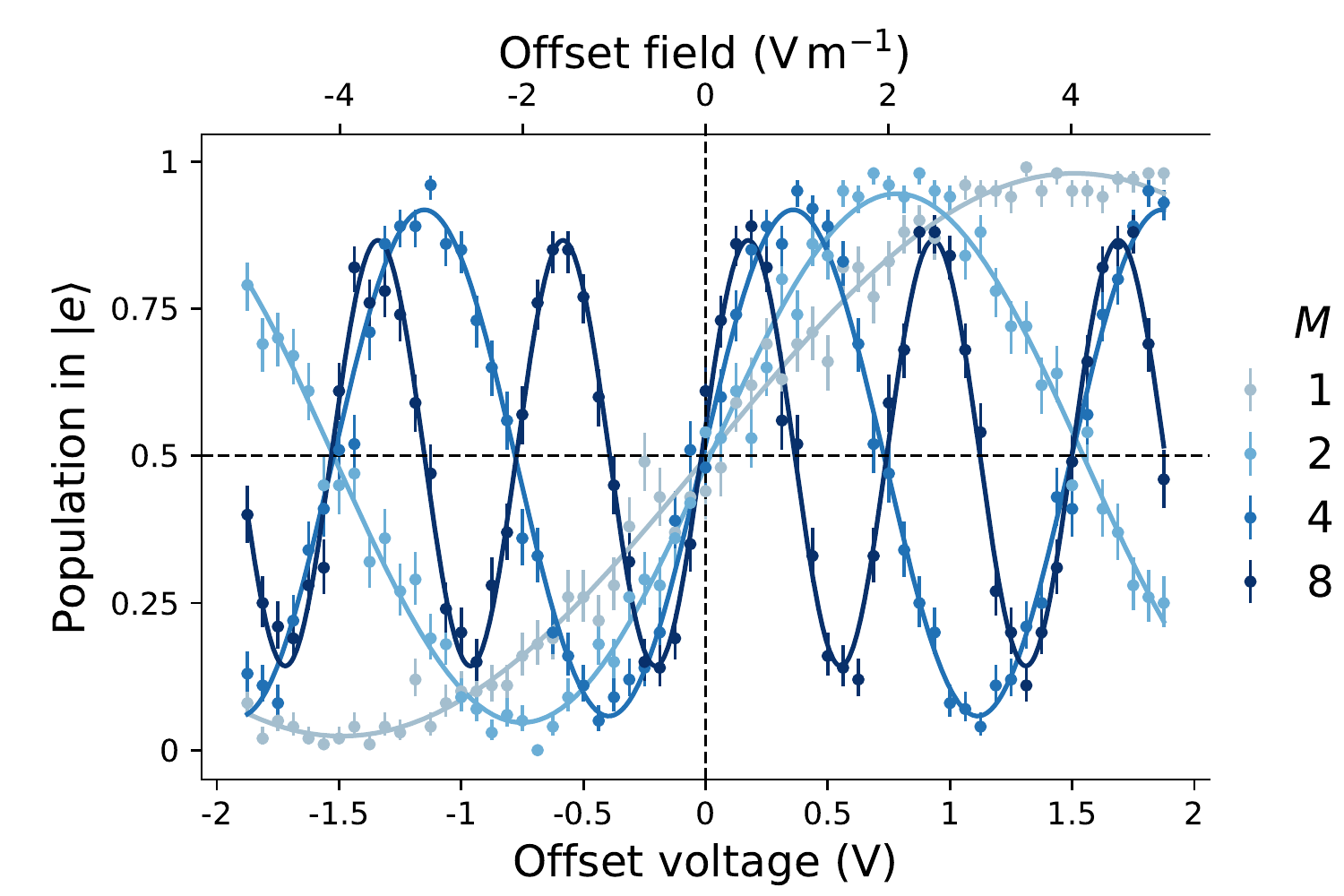}
\caption{Method~A becomes more sensitive to the compensation electrode offset voltage and to the offset field $\vec{E}$ with increasing sequence length $M$.
Solid lines represent sinusoidal fits to the data.
The oscillation contrast decreased as $M$ was increased due to the short coherence time of our system.
The $M=2$ dataset has a negative gradient at zero offset voltage because it was measured with $\theta_\mathrm{T}=\tfrac{\pi}{2}$ while the other measurements used $\theta_\mathrm{T}=-\tfrac{\pi}{2}$; for better comparison of the datasets we inverted the x-axis of the $M=2$ dataset.
Error bars represent quantum projection noise (1$\sigma$ confidence interval).
}
\label{fig_seq_multi_pulse_data}
\end{figure}

The probability $p$ of measuring the ion in $|e\rangle$ becomes more sensitive to $\phi_\mathrm{PD}$, and thus to a component of $\vec{E}$, as the sequence length $M$ is increased.
To show this we measured the dependence of $p$ on the compensation electrode offset voltage with sequences of different lengths $M$; the results are shown in Fig.~\ref{fig_seq_multi_pulse_data}.
The oscillation contrast decreased with increasing $M$, due to the limited coherence time in our experiment ($\sim 500\,\mathrm{\mu s}$ \cite{Lindberg2020}).

\subsection*{Method~B: Sequence using a fixed trap stiffness}
\begin{figure}[ht]
\centering
\includegraphics[width=\columnwidth]{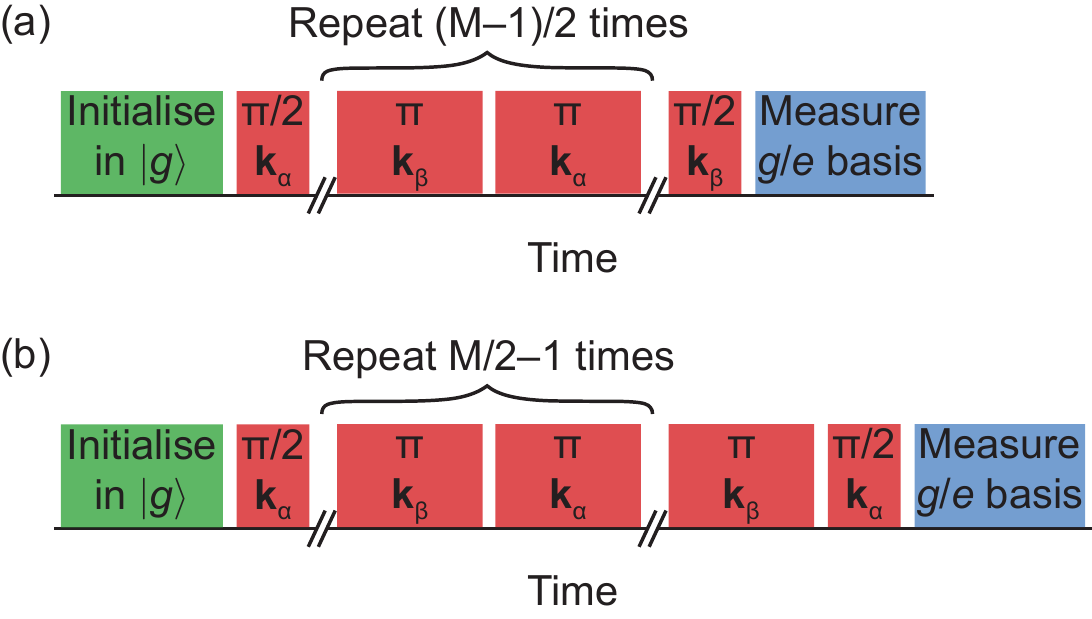}
\caption{Method~B involves coherent pulse sequences in which alternate pulses are driven by two different laser beams (with wavevectors $\vec{k}_\alpha$ and $\vec{k}_\beta$), while the trap stiffness is fixed.
The probability of measuring the ion in $|e\rangle$ reveals the phase difference between the laser fields at the ion position.
We measure the phase differences $\phi_\mathrm{PD}^A$ and $\phi_\mathrm{PD}^B$ at the ion equilibrium positions $\vec{r}_A$ and $\vec{r}_B$ when two different trap stiffness settings ($A$ and $B$) are used.
The quantity $\phi_\mathrm{PD}^A - \phi_\mathrm{PD}^B$ depends on $\vec{r}_{AB}$ and $\vec{E}$.
The sequences in (a) and (b) are used when $M$ is odd and even respectively.
The shortest sequence is a Ramsey sequence with $M=1$.
}
\label{fig_method_B}
\end{figure}

In the sequence described in this section the trap stiffness is fixed while the coherent pulses are applied, and alternate pulses are driven by two different laser beams.
This is represented in Fig.~\ref{fig_method_B}.

If the ion is at position $\vec{r}_A$ and alternate pulses are driven by two different laser beams, with wavevectors $\vec{k}_\alpha$ and $\vec{k}_\beta$, the phase $\phi_j$ alternates between two values $\Phi_{\alpha A}$ and $\Phi_{\beta A}$.
Using Eq.~(\ref{eq_Phi_alpha_A}), the difference between these phase values is
\begin{equation}
\Phi_{\alpha A}-\Phi_{\beta A} = \left( \vec{k}_\alpha - \vec{k}_\beta \right) \cdot \vec{r}_A + \Phi_{\alpha 0} - \Phi_{\beta 0}
\end{equation}
If the two laser beams are derived from the same source, the phase difference depends on the path length difference from the point where the beams are split to the ion position $\vec{r}_A$.
For convenience, we define $\phi_\mathrm{PD}^A \equiv \Phi_{\alpha A}-\Phi_{\beta A}$.
From Eq.~(\ref{eq_phi_T})
\begin{align}
\begin{split}
\phi_\mathrm{T} &= M \left( \Phi_{\alpha A}-\Phi_{\beta A} \right) \\
\phi_\mathrm{T} &= M \phi_\mathrm{PD}^A
\end{split}
\end{align}

If the sequence is conducted using the fixed trap stiffness $B$ then
\begin{align}
\begin{split}
\phi_\mathrm{T} &= M \left( \Phi_{\alpha B}-\Phi_{\beta B} \right) \\
&= M \left[ \left( \vec{k}_\alpha - \vec{k}_\beta \right) \cdot \vec{r}_B + \Phi_{\alpha 0} - \Phi_{\beta 0} \right] \\
&= M \phi_\mathrm{PD}^B
\end{split}
\end{align}
where $\phi_\mathrm{PD}^B \equiv \Phi_{\alpha B}-\Phi_{\beta B}$.
By conducting the sequence using each of the two trap stiffness settings, the phases $\phi_\mathrm{PD}^A$ and $\phi_\mathrm{PD}^B$ can be estimated, and therefrom the quantity
\begin{align}
\begin{split} \label{eq_phi_PD_AB}
\phi_\mathrm{PD}^A -& \phi_\mathrm{PD}^B = \left( \vec{k}_\alpha - \vec{k}_\beta \right) \cdot \left( \vec{r}_A - \vec{r}_B \right) \\
&= M \sum_i \frac{q (k_{\alpha i}-k_{\beta i}) E_i}{m} \left( \frac{1}{{\omega_{Ai}}^2} - \frac{1}{{\omega_{Bi}}^2} \right)
\end{split}
\end{align}
where, in the second line, Eq.~(\ref{eq_del_u}) is used.
$\phi_{\mathrm{PD}}^A - \phi_{\mathrm{PD}}^B$ reveals the difference between the ion equilibrium positions $\vec{r}_{AB}$ along the direction $\vec{k}_\alpha-\vec{k}_\beta$.
Thus, $\phi_\mathrm{PD}^A - \phi_\mathrm{PD}^B$ is sensitive to $\vec{E}$ along the direction $\vec{d}$ which has components
\begin{equation} \label{eq_d2}
d_i = (k_{\alpha i}-k_{\beta i}) \left( \frac{1}{{\omega_{Ai}}^2} - \frac{1}{{\omega_{Bi}}^2} \right)
\end{equation}

\begin{figure}[ht]
\centering
\includegraphics[width=\columnwidth]{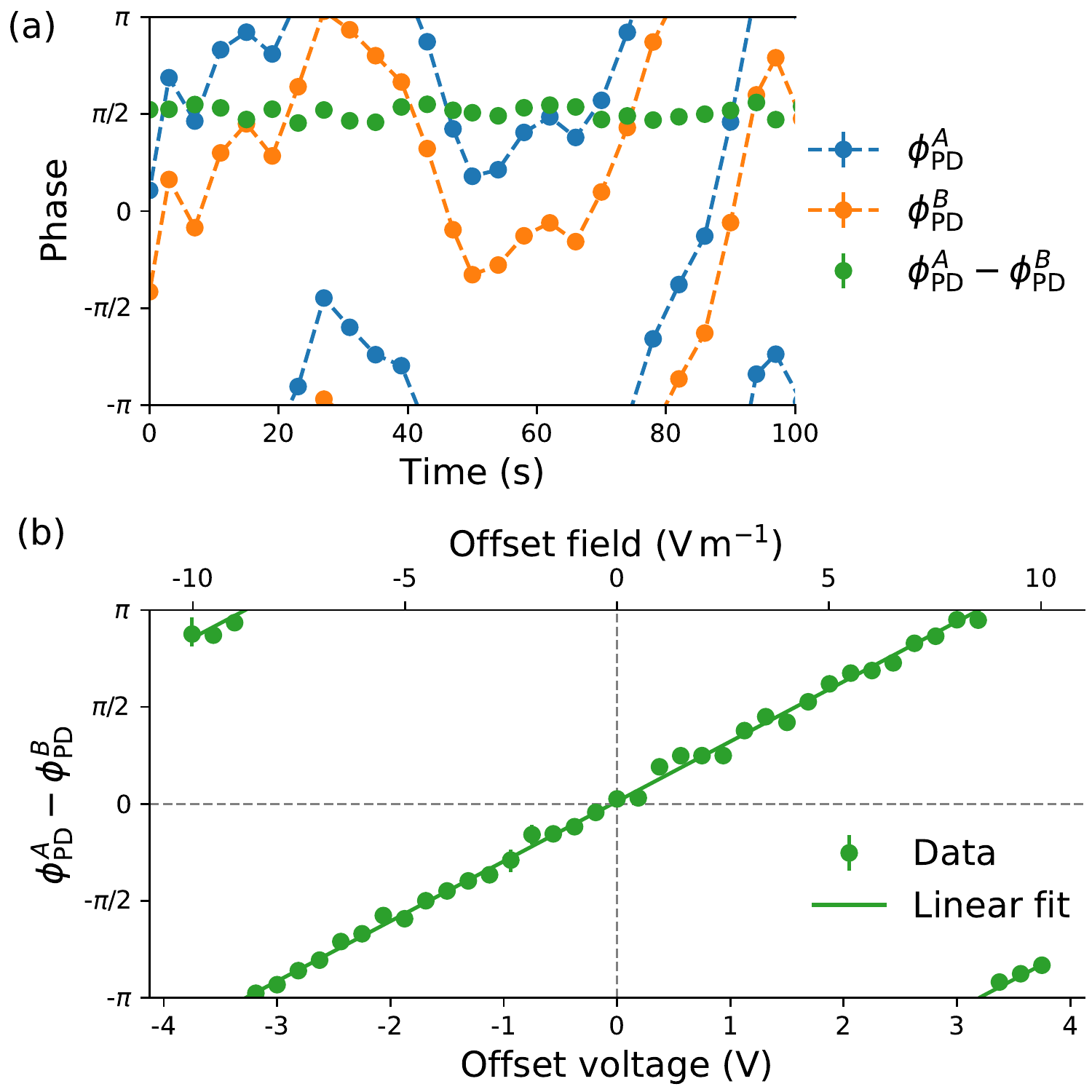}
\caption{Demonstration of Method~B.
(a)~$\phi_\mathrm{PD}^A$ and $\phi_\mathrm{PD}^B$ were repeatedly measured over 100\,s.
The phase difference $\phi_\mathrm{PD}^A-\phi_\mathrm{PD}^B$ is stable in time, despite the limited interferometric stability between the two beams, which causes $\phi_\mathrm{PD}^A$ and $\phi_\mathrm{PD}^B$ to drift.
(b)~$\phi_\mathrm{PD}^A-\phi_\mathrm{PD}^B$ has a linear dependence on the offset voltage applied to a micromotion compensation electrode.
Error bars representing quantum projection noise (1$\sigma$ confidence interval) are generally smaller than the marker size.
}
\label{fig_PD_A_B}
\end{figure}

We demonstrated this method in our system, the results are shown in Fig.~\ref{fig_PD_A_B}.
The two laser beams are derived from the same source, they are each passed through a separate acousto-optic modulator (allowing each beam to be separately switched on and off, and allowing controlled phase shifts $\{\theta_j\}$ to be introduced), then each beam is guided through an optical fiber before it is focussed onto the ion.
The path length difference from the point where the beams are separated to the experimental chamber, varies in time, due to temperature fluctuations and mechanical vibrations.
Because of this $\left( \Phi_{\alpha 0} - \Phi_{\beta 0} \right)$ and thus $\phi_{\mathrm{PD}}^A$ and $\phi_{\mathrm{PD}}^B$ vary in time.
We measured the drift of $\phi_{\mathrm{PD}}^A$ and $\phi_{\mathrm{PD}}^B$ in time using the sequence of Fig.~\ref{fig_method_B} with $M=1$ by interleaving measurements using trap settings $A$ and $B$, and control phases $\theta_\mathrm{T}=0$ and $-\tfrac{\pi}{2}$, and using Eq.~(\ref{eq_phiT_3}).
The results are shown in Fig.~\ref{fig_PD_A_B}(a).

Because $\phi_{\mathrm{PD}}^A$ and $\phi_{\mathrm{PD}}^B$ do not drift too fast, and because the difference between the ion equilibrium positions $\vec{r}_{AB}$ is stable, the difference between the estimates $\phi_{\mathrm{PD}}^A-\phi_{\mathrm{PD}}^B$ is stable in time, as shown in Fig.~\ref{fig_PD_A_B}(a).

We varied the voltage applied to a compensation electrode and measured the linear response of $\phi_{\mathrm{PD}}^A - \phi_{\mathrm{PD}}^B$, this is shown in Fig.~\ref{fig_PD_A_B}(b).
This result is consistent with Eq.~(\ref{eq_phi_PD_AB}), which describes a linear relationship between $\phi_{\mathrm{PD}}^A - \phi_{\mathrm{PD}}^B$ and a component of $\vec{E}$.
Thus, the quantity $\phi_{\mathrm{PD}}^A - \phi_{\mathrm{PD}}^B$ can be used for minimizing micromotion.

Longer pulse sequences (with larger $M$) offer more precise measurement of $\phi_{\mathrm{PD}}^A - \phi_{\mathrm{PD}}^B$, though they also require better interferometric stability between the two beams.
This can be achieved using active stabilisation \cite{Ma1994}.

\section{Fast and accurate micromotion minimization} \label{sec_fast_accurate}
\begin{figure}[ht]
\centering
\includegraphics[width=\columnwidth]{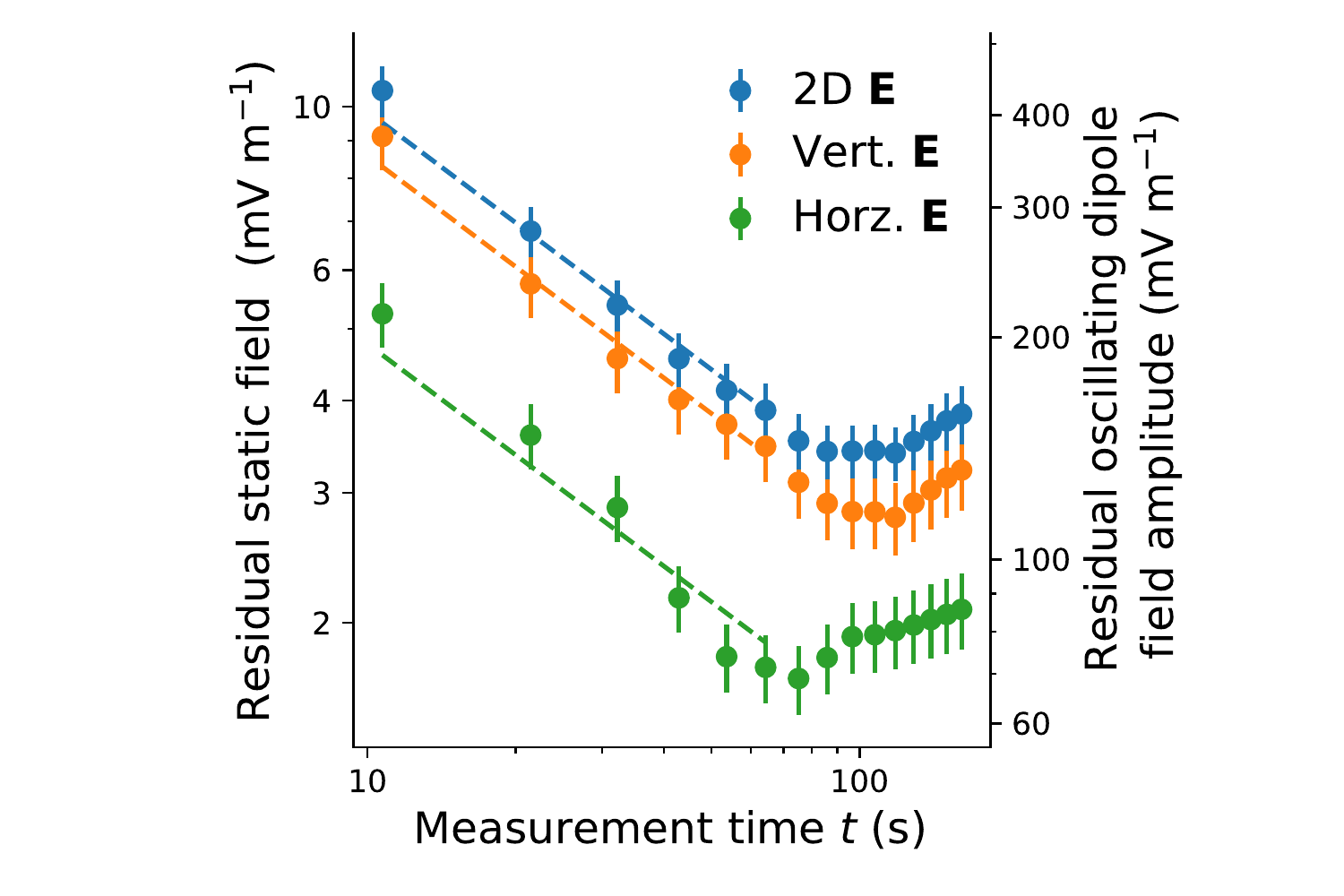}
\caption{Using Method~A we minimized the magnitude of the offset field $\vec{E}$ quickly and accurately.
With increasing measurement time $t$ the residual field strength decreased as $t^{-1/2}$, until around 100\,s when drifts caused the accuracy to worsen.
Dashed lines are $t^{-1/2}$ fits.
Error bars represent the standard error of the mean (1$\sigma$ confidence interval).
}
\label{fig_allan}
\end{figure}

Using Method~A with $M=8$ we minimized the strength of the offset field $\vec{E}$ quickly and accurately.
This is shown by the data in Fig.~\ref{fig_allan}.
The experiment runs alternated between using two different laser beams from two different directions; in this way we probed $\vec{E}$ in two dimensions, i.e.\ the plane of the oscillating field of a linear Paul trap.
We see that with increasing measurement time $t$ the residual electric field strength decreased as $t^{-1/2}$, until around 100\,s when drifts kicked in.
The drifts were likely caused by changes in the offset field $\vec{E}$ and instability of the voltage sources used to apply voltages to the compensation electrodes.

We obtained the data as follows:
First we measured the rate of change of $\phi_\mathrm{PD}$ with respect to compensation electrode voltage, in much the same way as shown in Fig.~\ref{fig_seq1}(b).
We did this for $\phi_\mathrm{PD}$ measurements using the two different laser beams and two different compensation electrodes.
Then we repetitively measured $\phi_\mathrm{PD}$ using the two different laser beams, and every 11\,s we updated the voltages of the two compensation electrodes so as to minimize $|\vec{E}|$ in two dimensions.
We measured repetitively over 18~minutes.
By analysing data collected over this time, we see how well the magnitude of the electric field $\vec{E}$ can be minimised with different measurement times.
The analysis is much the same as that used to calculate the overlapping Allan deviation of fractional frequency data from a clock.

After 75\,s of measurement the 2D residual static field strength was $(3.5\pm0.3)\,\mathrm{mV\,m^{-1}}$, which is, as far as we are aware, lower than the residual static field strength achieved using any other micromotion minimization technique, and also lower than the residual field achieved in a system of optically-trapped ions \cite{Huber2014}.
The field uncertainty decreased with increasing measurement time as $(31.1\pm1.0)\,\mathrm{mV\,m^{-1}\,Hz^{-1/2}}$.
The horizontal component of $\vec{E}$ was minimised faster than the vertical component, since the beam probing the horizontal component has a larger projection onto the plane of the oscillating field than the beam probing the vertical component.

On the second y-axis of Fig.~\ref{fig_allan} we show the corresponding strength of the residual oscillating dipole field experienced by the ion, which arises because the offset field $\vec{E}$ displaces the ion from the oscillating quadrupole field null.
This assumes that there is no additional oscillating dipole field in our system which arises from a phase mismatch of the voltages applied to the trap electrodes (quadrature micromotion) \cite{Berkeland1998}.
It's worth noting that a horizontal (vertical) offset field $\vec{E}$ causes the ion to experience a vertical (horizontal) oscillating dipole field (see Fig.~\ref{fig_sb_method}).

The experiments were evenly split between using two different laser beams and four different sets of control phases $\{\theta_j\}$.
The use of four sets of controllable phases diminished systematic errors, as is described in Section~\ref{sec_dynamical_decoupling}.
During these measurements we reduced the ion's radial secular frequencies from $2\pi\times1.5\,\mathrm{MHz}$ to $2\pi\times840\,\mathrm{kHz}$, while keeping the axial secular frequency at $2\pi\times350\,\mathrm{kHz}$.
The oscillating quadrupole field's frequency was $2\pi \times 18.1\,\mathrm{MHz}$.

Faster minimization of $\vec{E}$ could be achieved using a larger change of the trap stiffness or by using a longer sequence (with higher $M$).
A phase estimation sequence with over 1000 pulses has been conducted in an experimental setup with a longer coherence time than ours \cite{Rudinger2017}.
With such long sequences care must be taken to mitigate heating of the ion's motion caused by the trap stiffness changes.
Ion heating causes pulse area errors, which in turn can cause systematic errors in $\phi_\mathrm{PD}$ estimates.
This can be mitigated by changing the trap stiffness sufficiently slowly, or by employing sympathetic cooling \cite{Barrett2003, Home2009}.
Alternatively $\vec{E}$ can be probed using Method B, which does not involve changes of the trap stiffness between the pulses.

\subsection*{Changing the RF power applied to the trap electrodes affects the trap temperature}

After just $11\,\mathrm{s}$ measurement time we achieve a low residual oscillating dipole field, which would cause a second-order Doppler shift on the $\mathrm{^{88}Sr^+}$ clock transition below the $10^{-22}$ level \cite{Berkeland1998, Keller2015}.
And so, the micromotion minimization methods presented here stand to benefit precision spectroscopy experiments.
However, in precision spectroscopy experiments, care should be taken to mitigate unwanted changes of the trap temperature:

Changing the trap stiffness by changing the RF power supplied to the trap electrodes affects the RF power dissipated in the system, which, in turn, affects the trap temperature.
Changes of the trap temperature affect the blackbody radiation field experienced by the ion.
Further, thermal expansion can shift the relative positions of trap electrodes and affect $\vec{E}$ \cite{Gloger2015}, and also cause beam-pointing errors.
During the measurements used to produce the data shown in Fig.~\ref{fig_allan} we did not make efforts to mitigate trap temperature changes.
During these measurements the RF signal applied to the trap electrodes was reduced 4\% of the time.
We estimate that the decrease of the average RF power caused the temperature of the ion's surroundings to decrease by $\sim 10\,\mathrm{mK}$ \cite{Guggemos2017}, causing a blackbody radiation shift on the $\mathrm{^{88}Sr^+}$ clock transition $\sim 10^{-19}$ \cite{Dube2014}.

\begin{figure}[ht]
\centering
\includegraphics[width=\columnwidth]{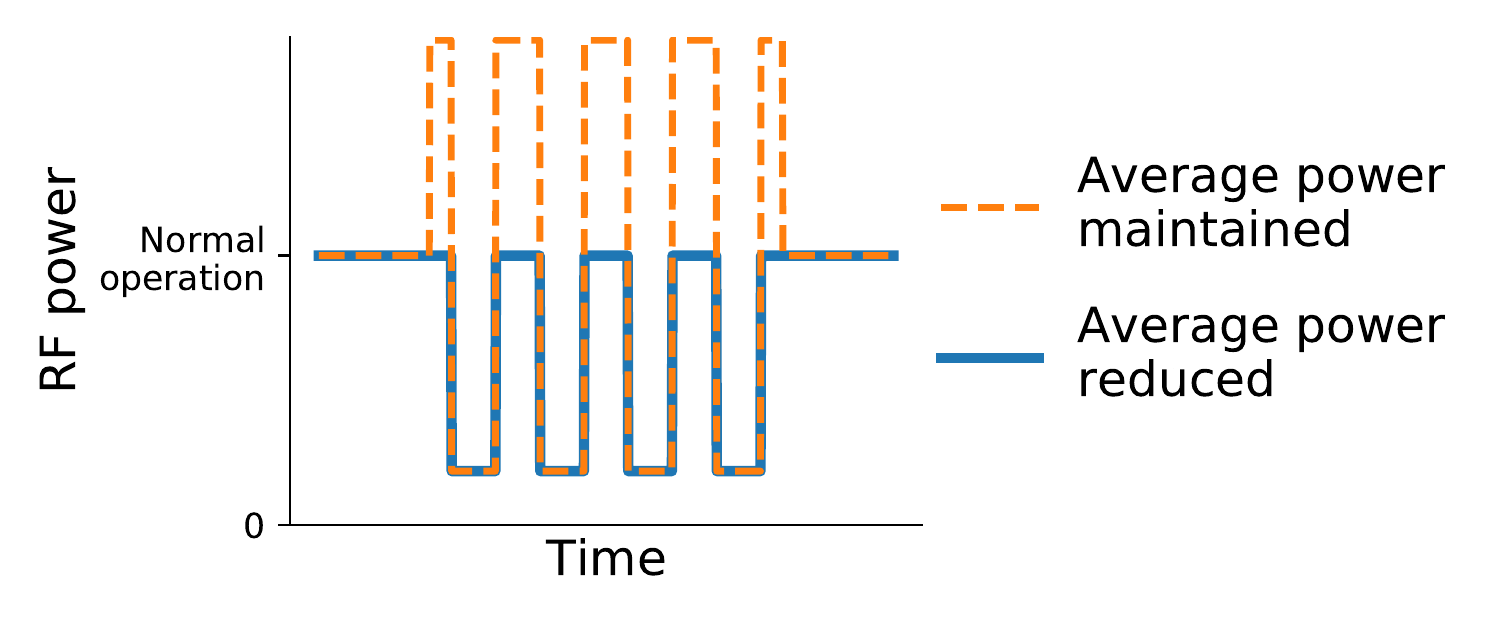}
\caption{To mitigate unwanted changes of the trap temperature, the average RF power used during the interferometry sequences should equal the RF power used during normal trap operation.
The RF power profiles sketched here are suitable for use with Method~A and $M=8$.
}
\label{fig_trap_temp_schematic}
\end{figure}

To mitigate trap temperature changes the average RF power used during the micromotion minimization sequences should equal the RF power used during the trap's normal operation \cite{Gloger2015}, for instance as sketched in Fig.~\ref{fig_trap_temp_schematic}.
Alternatively, the trap stiffness can be changed during the micromotion minimization sequences by changing the amplitude of the trap's static quadrupole field \cite{Schneider2005, Gloger2015}.

\section{Micromotion minimization with sub-standard quantum limit scaling using a binary search algorithm} \label{sec_efficient_phase_estimation}
In this section we use a binary search algorithm (based on the \textit{Robust Phase Estimation} technique \cite{Kimmel2015}) to efficiently measure $\phi_\mathrm{PD}$ of Method~A with an uncertainty below the standard quantum limit (SQL).
The same methodology can be used together with Methods B or C (Appendix~\ref{appendix_method_c}).
This phase estimation technique can be used to achieve Heisenberg scaling, it is easy to implement, the data analysis is straightforward and the protocol is non-adaptive.
While adaptive phase estimation techniques allow for more accurate phase measurements at the Heisenberg limit \cite{Boixo2008, Higgins2007}, they require measurement settings to be updated on the fly, which is not possible with our current control system \cite{Pham2005, Schindler2008, Heinrich2020}.

\begin{figure}[ht]
\centering
\includegraphics[width=\columnwidth]{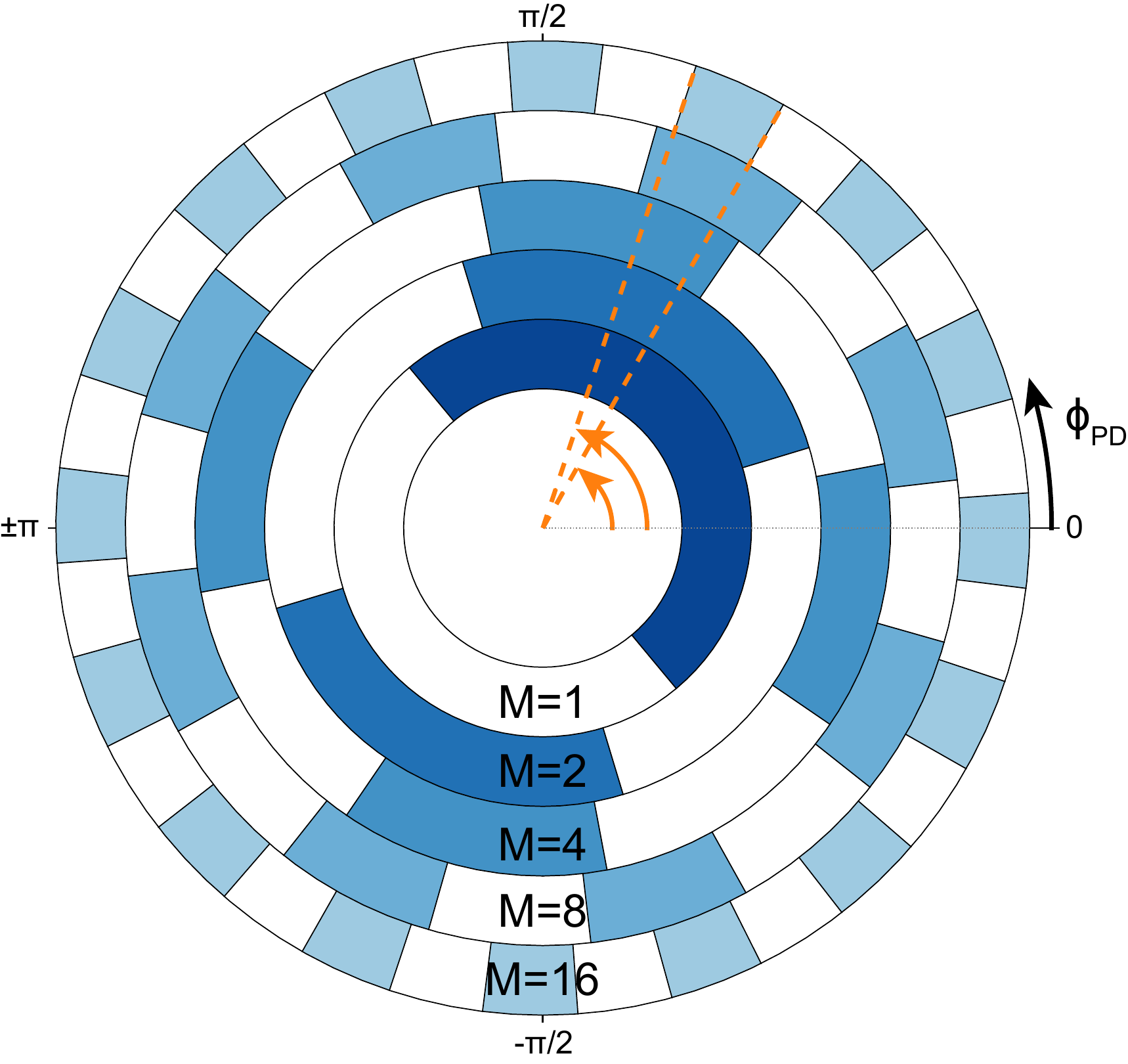}
\caption{Illustration of the binary search algorithm (based on the \textit{Robust Phase Estimation} technique \cite{Kimmel2015}).
By conducting measurements using different sequence lengths $\phi_{\mathrm{PD}}$ can be determined efficiently.
The shaded regions of width $\pi/M$ indicate values of $\phi_{\mathrm{PD}}$ consistent with the measurement results.
Shorter sequences allow $\phi_{\mathrm{PD}}$ to be reckoned from within a larger range, but they are less precise.
Longer sequences are more precise, but they allow $\phi_{\mathrm{PD}}$ to be reckoned from within only a narrow range.
By combining the results $\phi_{\mathrm{PD}}$ can be determined with high precision from a broad range.
The orange arrows indicate the range of $\phi_{\mathrm{PD}}$ values consistent with all the measurement results.
}
\label{fig_RPE_explanation}
\end{figure}

The binary search algorithm works as follows:
Starting with an unknown phase $\phi_\mathrm{PD}$ from within the range $[-\pi, \pi]$, a set of measurements are first conducted using a sequence with $M=1$ to limit $\phi_\mathrm{PD}$ to a range of width $\pi$, then a set of measurements with $M=2$ narrow the range to $\pi/2$, then a set of measurements with $M=4$ narrow the range to $\pi/4$, and so on.
The $j^\mathrm{th}$ set of measurements use a sequence with $M_j=2^{j-1}$ to narrow the range to a width of $\pi/2^{j-1}$.
The technique is illustrated in Fig.~\ref{fig_RPE_explanation}.

\begin{figure}[ht]
\centering
\includegraphics[width=\columnwidth]{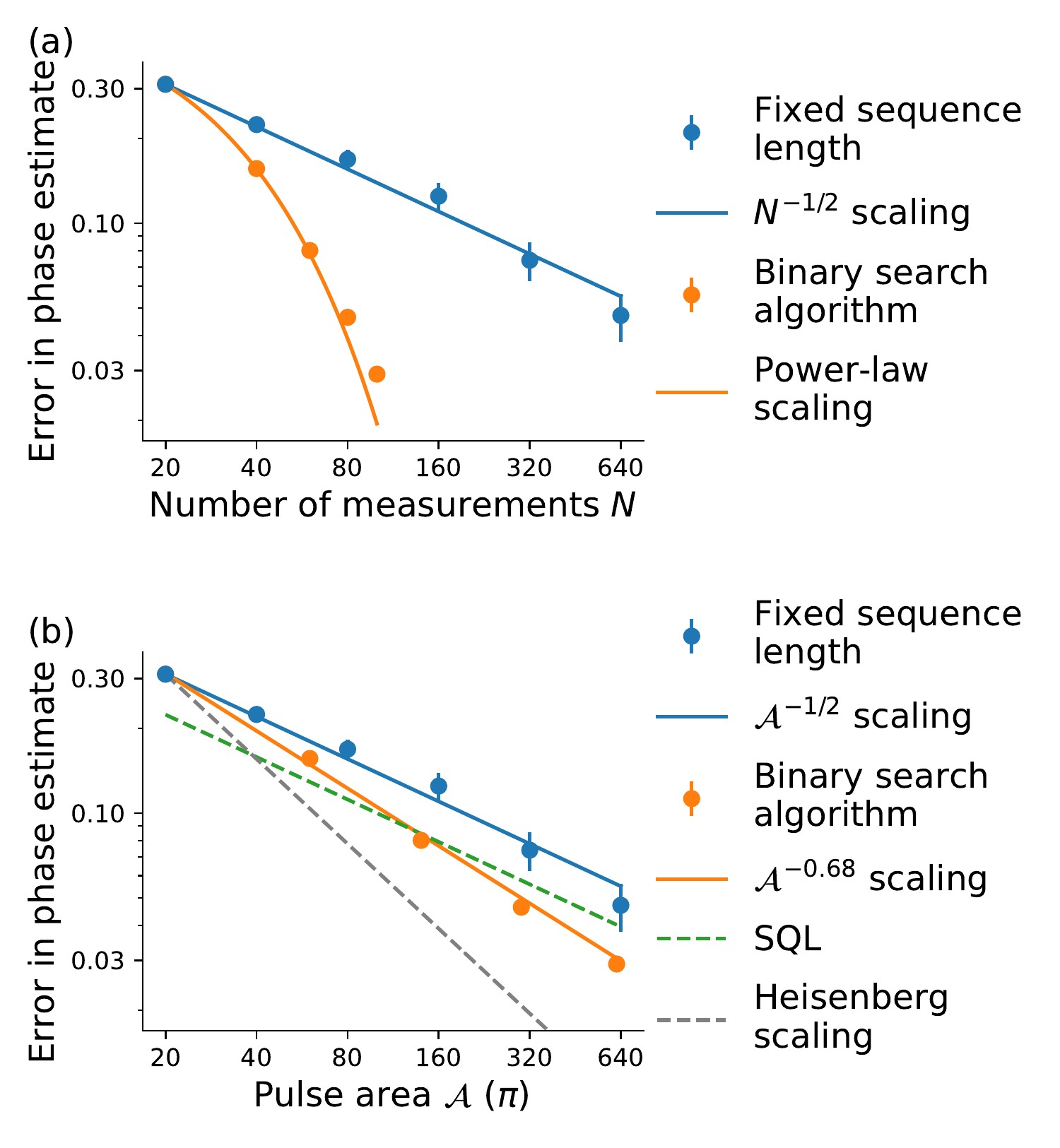}
\caption{$\phi_\mathrm{PD}$ can be efficiently measured using a binary search algorithm.
Combining the results of measurements using different sequence lengths (orange data) is more efficient than using a fixed sequence length (blue data).
This is true both when the number of measurements conducted is considered, as in (a), and when the total pulse area is considered, as in (b).
Using the binary search algorithm a $\phi_\mathrm{PD}$-uncertainty lower than the standard quantum limit (SQL) is achieved.
Error bars represent the standard error of the mean (1$\sigma$ confidence interval).
}
\label{fig_rpe}
\end{figure}

We demonstrated the efficiency of this protocol as follows:
We carried out 59,000 measurement runs, split evenly between five sequence lengths $M \in \{1,2,4,8,16\}$.
The measurements were also split between using two different $\theta_\mathrm{T}$ values.
We then analysed the estimates of $\phi_\mathrm{PD}$ given by sub-sampled datasets.
If we consider first the results using only $M=1$ data, the error in the estimates of $\phi_\mathrm{PD}$ decreased with the number of measurements in the sample $N$ as $N^{-1/2}$.
This is shown by the blue data in Fig.~\ref{fig_rpe}(a).
The binary search algorithm allows improved estimates to be achieved using fewer measurements, as shown by the orange data.
The first orange datapoint describes the error in estimates of $\phi_\mathrm{PD}$ using 40 measurements split evenly between different sequence lengths $M\in\{1,2\}$.
The second orange datapoint describes the error in estimates using 60 measurements split evenly between sequence lengths $M\in\{1,2,4\}$.
And so on for the third and fourth orange datapoints.
The scaling of the orange data with the number of measurements is well described by a power law.
The deviation from the power law for the final datapoint (using sequences with up to $M=16$) is due to the limited coherence time of our experiment and also because of the non-zero probability of error in the results of the measurements with $M<16$, which contribute to the overall estimate.
The ``true'' value of $\phi_\mathrm{PD}$ was estimated using all 59,000 measurements.

The duration of each experimental run was dominated by cooling and fluorescence detection, rather than the duration of the coherent pulses.
Thus, the x-axis in Fig.~\ref{fig_rpe}(a) reflects the total measurement time.
For long sequences with large $M$ the total measurement time would be better represented by the total area of coherent pulses $\mathcal{A}$ than by the number of measurements \cite{Higgins2009}.
And so we rescale the x-axis of Fig.~\ref{fig_rpe}(a) to view the scaling of the same data with the pulse area, this is shown in Fig.~\ref{fig_rpe}(b).
Here we see that the binary search algorithm allows us to estimate $\phi_\mathrm{PD}$ with an error below the SQL $\sqrt{\frac{\pi}{\mathcal{A}}}$ \cite{Giovannetti2004}.
A better scaling would be achieved in an experimental setup with a longer coherence time.
Also, to achieve Heisenberg scaling the different measurement sets (parameterised by $j$) need to use different numbers of measurements $N_j$ \cite{Kimmel2015}.

For readers interested in using the binary search algorithm in their systems we reproduce an algorithm for combining the results of different measurement sets from \cite{Rudinger2017} in Appendix~\ref{appendix_rpe_algorithm}.

\section{Robust estimates of \texorpdfstring{$\phi_\mathrm{T}$}{phases} using suitable control phases \texorpdfstring{$\{\theta_j\}$}{}} \label{sec_dynamical_decoupling}
Changing the overall control phases $\theta_T$ [Eq.~(\ref{eq_theta_T})] shifts the dependence of $p$ on $\phi_T$, as can be appreciated from Eq.~(\ref{eq_pe_cos_theta_T_phi_T}) and Fig.~\ref{fig_seq1}(a).
By appropriately choosing the individual phases $\theta_j$, estimates of $\phi_T$ can be made robust against laser detuning and pulse area errors.
Pulse area errors can arise if the sequence includes fast changes of the trap stiffness, which can cause motional heating, which in turn modifies the $|g\rangle \leftrightarrow |e\rangle$ coupling strength.
They can also be caused by the change in laser light intensity when the ion changes position within a tightly-focussed laser beam.

We used simulations to test different sets of control phases $\{\theta_j\}$ when the pulse sequences from Method~A and Method~B are used, and we found $\phi_\mathrm{T}$ (and thus $\phi_\mathrm{PD}$, $\phi_\mathrm{PD}^A$ and $\phi_\mathrm{PD}^B$) can be robustly estimated in the presence of these errors using
\begin{align}
\text{Settings I:   \quad} &\theta_j =\begin{cases}
0 &\text{even $j$} \\
-\tfrac{\pi}{2} &\text{odd $j$, } 1<j<M+1 \\
\pi &j=M+1
\end{cases} \\
\text{Settings II:  \quad} &\theta_j=\begin{cases}
0 &\text{even $j$} \\
\tfrac{\pi}{2} &\text{odd $j$, } 1<j<M+1 \\
\pi &j=M+1
\end{cases}
\end{align}
and $\theta_1 \in \{\pi, \tfrac{\pi}{2}\}$, and where $M$ is an even integer and $\phi_\mathrm{T}$ is small.
Using these settings $\phi_\mathrm{T}$ can be estimated using
\begin{align}
\begin{split}
\phi_{\mathrm{T}} = \mathrm{arctan2} \Big\{ &  (-1)^{M/2} \left[ p{\left( \theta_1=\tfrac{\pi}{2} \right)} - \tfrac{1}{2} \right], \\
& (-1)^{M/2} \left[ p{\left( \theta_1=\pi \right)} - \tfrac{1}{2} \right]   \Big\}
\end{split}
\end{align}

\begin{figure}[ht]
\centering
\includegraphics[width=0.9\columnwidth]{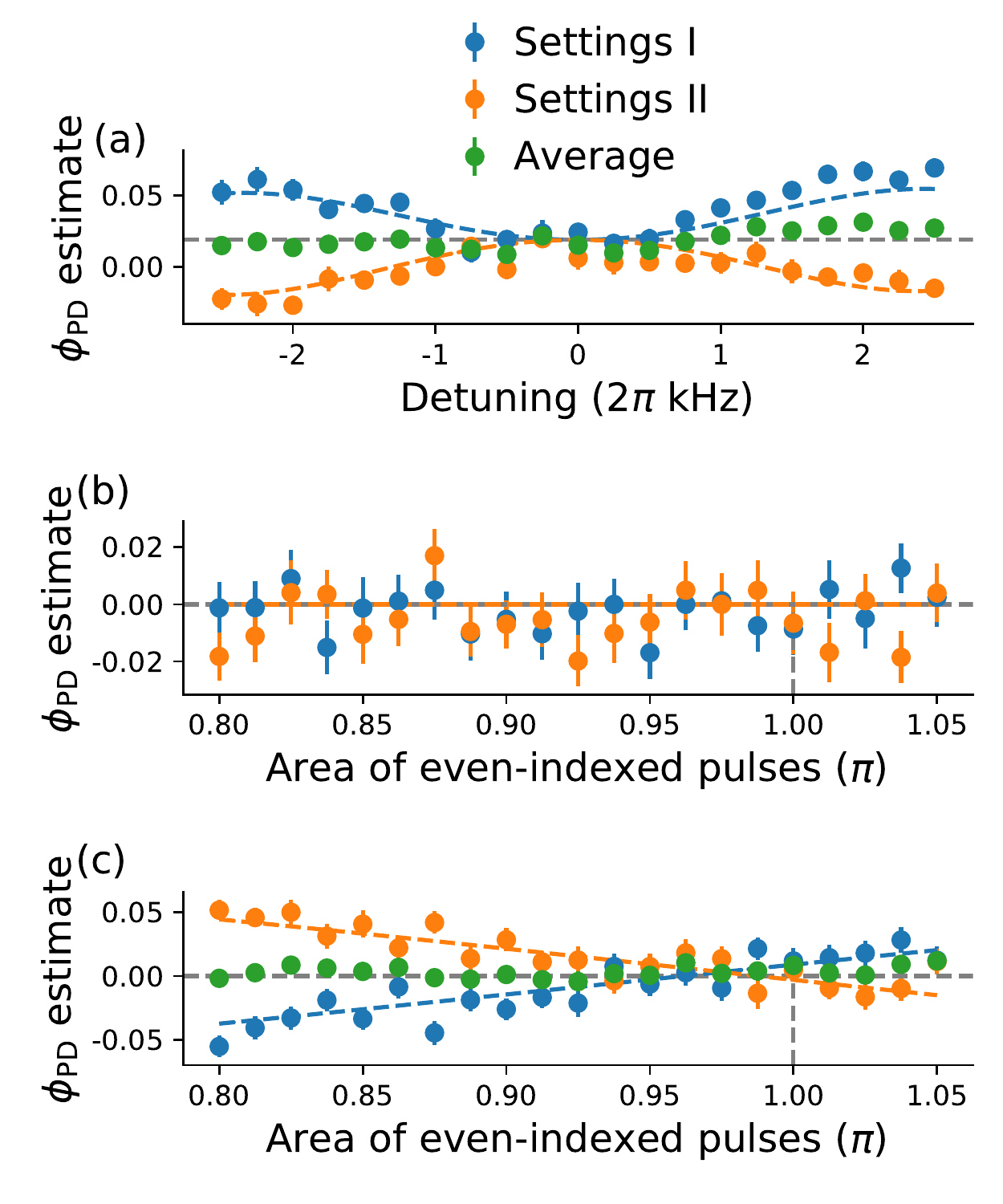}
\caption{Estimates of $\phi_\mathrm{PD}$ (and $\phi_\mathrm{PD}^A$ and $\phi_\mathrm{PD}^B$) are robust against errors when the control phase $\{\theta_j\}$ settings I and II are used.
The robustness is improved by averaging the estimates obtained using settings I and II.
(a)~The detuning of the laser from the $|g\rangle \leftrightarrow |e\rangle$ resonance was scanned.
(b)-(c)~A pulse area error on even-indexed pulses was introduced and varied, in (b) no additional errors were added, in (c) a 10\% error was introduced to the area of odd-indexed pulses.
Dashed lines indicate simulation results, which show good agreement with the experimental results; the only free parameter was a phase offset used in (a) which accounts for a weak offset field $\vec{E}$.
Error bars represent quantum projection noise (1$\sigma$ confidence interval).
}
\label{fig_robustness}
\end{figure}

We experimentally tested the robustness of $\phi_\mathrm{PD}$ estimates by introducing different errors to our system.
The results are shown in Fig.~\ref{fig_robustness}.

First we measured $\phi_\mathrm{PD}$ of Method~A with $M=16$, when different laser detunings were used.
One might expect that a detuning $\Delta$ might shift $\phi_\mathrm{T}$ by $\Delta \cdot T$ and $\phi_\mathrm{PD}$ by $\Delta \cdot T/M$, where the duration of the coherent pulse sequence $T$ we used was $1.6\,\mathrm{ms}$.
The $\phi_\mathrm{PD}$ estimates using settings I and II were much more stable than this, as shown in Fig.~\ref{fig_robustness}(a).
Furthermore, the estimate of $\phi_\mathrm{PD}$ becomes still more stable by averaging the estimates obtained with settings I and II.

Then we investigated the robustness in the presence of pulse area errors.
We conducted experiments with a pulse area error on the even-indexed pulses.
The phase estimates were stable when the magnitude of this error was varied, as shown in Fig.~\ref{fig_robustness}(b).
We additionally introduced a 10\% pulse area error on the odd-indexed pulses, and found that the robustness of the phase estimate could again be improved by averaging the results of experiments conducted using settings I and settings II, as shown in Fig.~\ref{fig_robustness}(c).
These experiments used $M=8$, a fixed trap stiffness and a single laser beam driving the pulses.
The reason we alternated the pulse area error between pulses is that this will happen in practice, since in Method~A the trap stiffness setting is alternated, while in Method B the laser beam is alternated.

\begin{figure}[ht]
\centering
\includegraphics[width=\columnwidth]{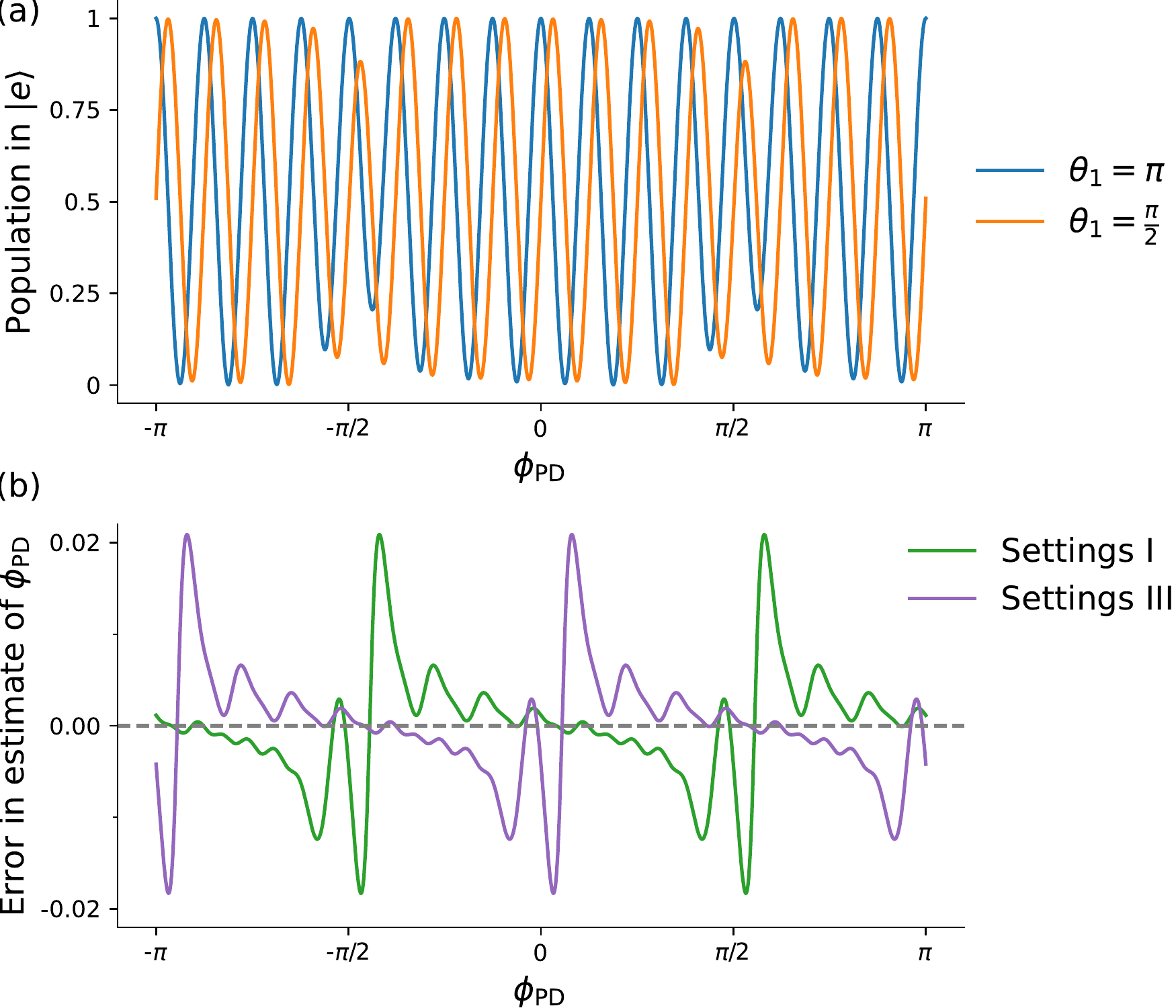}
\caption{Pulse area errors and laser detuning cause systematic errors in estimates of $\phi_\mathrm{PD}$, the size of the errors depend on the control phases $\{\theta_j\}$ used and the true value of $\phi_\mathrm{PD}$.
We show this using simulated experimental runs using sequences with $M=16$ in which the even-indexed pulses have a 5\% pulse area error.
(a) Control phase settings I were used and the pulse area error affected the probability $p$ of measuring the ion in state $|e\rangle$ most strongly around $\phi_\mathrm{PD}=\pm\tfrac{\pi}{2}$ where the oscillation contrast was reduced.
(b) Estimates of $\phi_\mathrm{PD}$ were generated from the simulated runs.
With control phase settings I the accuracy is highest near $\phi_\mathrm{PD}=0$ or $\pi$, while with settings III the accuracy is highest near $\phi_\mathrm{PD}=\pm \pi/2$.
}
\label{fig_robustness_simulation}
\end{figure}

The robustness properties depend on the size of the phase $\phi_\mathrm{PD}$, as shown by the simulation results in Fig.~\ref{fig_robustness_simulation}.
We simulated experimental runs using pulse sequences of length $M=16$ with pulse area errors of 5\% on the even-indexed pulses.
The results of simulations using control phase settings I are shown in Fig.~\ref{fig_robustness_simulation}(a); we see that the probability $p$ of measuring the ion in state $|e\rangle$ deviates from the unity-contrast oscillations described by Eq.~(\ref{eq_pe_M_phiPD}), at around $\phi_\mathrm{PD}=\pm\tfrac{\pi}{2}$.
From this data estimates of $\phi_\mathrm{PD}$ were generated, and the systematic errors in the estimates (caused by the 5\% pulse area error) were largest when the true value of $\phi_\mathrm{PD}$ was around $\pm\tfrac{\pi}{2}$, as shown in Fig.~\ref{fig_robustness_simulation}(b).
We also simulated measurements using control phase settings III (described in Appendix \ref{appendix_settings_III}), then the systematic errors in $\phi_\mathrm{PD}$ estimates were largest when the true value of $\phi_\mathrm{PD}$ was near $0$ or $\pi$.

Although the robustness of phase estimates depends on the true value of the phase, this is unlikely to be a problem when Method~A is used and when micromotion is nearly minimized -- then $\phi_\mathrm{PD}$ is small and control phase settings I and II perform well.
However, if Method~B is used in an experiment setup in which the path length difference between the two laser beams is not stable, then $\phi_\mathrm{PD}^A$ and $\phi_\mathrm{PD}^B$ will drift over time [as shown in Fig.~\ref{fig_PD_A_B}(a)] and the robustness of the phase estimates will be unstable.
This instability could be mitigated by adapting the control phase values $\{\theta_j\}$ during a measurement, or by actively stabilising the path length difference between the two beams \cite{Ma1994}.

\section{Micromotion minimization in 2D and 3D} \label{sec_2D_and_3D_main}

\subsection{Applying the interferometry methods in 2D and 3D}
To counter an unwanted electric field $\vec{E}$ in 2D (3D) we produce a 2D (3D) compensating field by supplying voltages to two (three) compensation electrodes.
To determine the appropriate voltages we need to measure $\phi_\mathrm{PD}$ or $\phi_\mathrm{PD}^A-\phi_\mathrm{PD}^B$ using two (three) laser beam configurations.
First we measure the dependence of the $i^\mathrm{th}$ phase measurement on the $j^\mathrm{th}$ compensation electrode voltage, in the same way as in Fig.~\ref{fig_seq1}(b) or Fig.~\ref{fig_PD_A_B}(b).
We label the gradient of this dependence $\mathcal{M}_{ij}$.
We use the four (nine) $\mathcal{M}_{ij}$ values to construct a 2$\times$2 (3$\times$3) matrix $\mathbfcal{M}$.
Then we can minimize $|\vec{E}|$ by measuring the two (three) phase values, storing them in a two-element (three-element) vector $\boldsymbol{\phi}$, then calculating $\vec{V} = \mathbfcal{M}^{-1} \cdot \boldsymbol{\phi}$ \cite{Roos2000}.
The two-element (three-element) vector $\vec{V}$ describes the offsets of the compensation electrode voltages from the optimal values.
Note that the matrix $\mathbfcal{M}$ depends on the trap settings used.

If one wishes to relate $\boldsymbol{\phi}$ to the offset field $\vec{E}$, one can use Eq.~(\ref{eq_phiT_E}) or Eq.~(\ref{eq_phi_PD_AB}).
This requires knowledge of the direction of the laser field wavevectors and the change of the secular frequencies.
Alternatively one can relate $\vec{V}$ to $\vec{E}$ using another micromotion minimization technique; in this work we related $\vec{V}$ to $\vec{E}$ using the resolved sideband method \cite{Berkeland1998, Keller2015}.

\subsection{2D micromotion minimization using a single probe laser beam} \label{sec_2d_compensation_one_beam}
\begin{figure}[ht]
\centering
\includegraphics[width=\columnwidth]{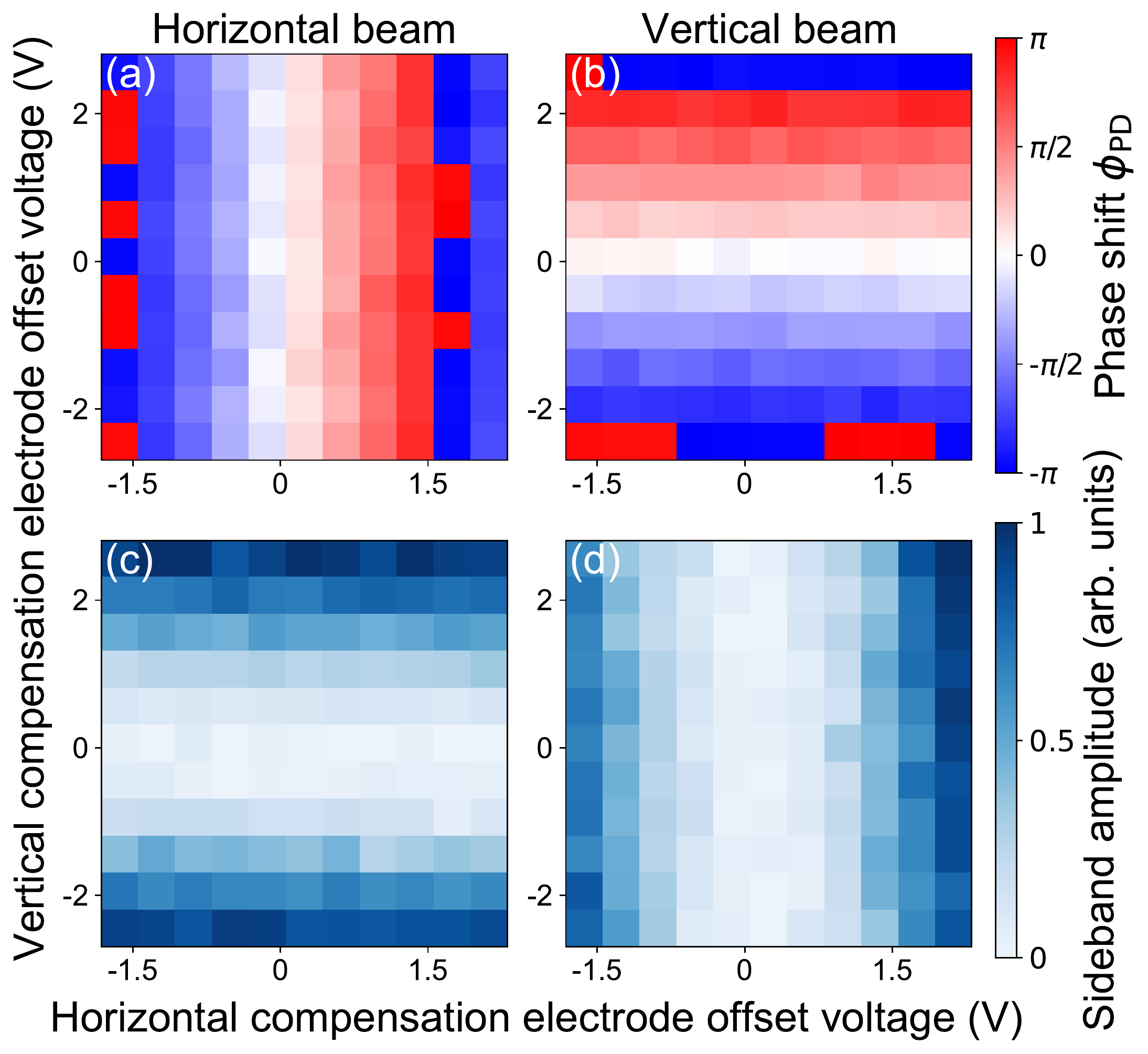}
\caption{$|\vec{E}|$ can be minimized in 2D using a single probe laser, by using interferometry method~A together with the resolved sideband technique \cite{Berkeland1998}.
(a) and (b) [(c) and (d)] use the interferometry (resolved sideband) method, (a) and (c) [(b) and (d)] use a horizontal (vertical) probe beam.
}
\label{fig_1_beam_2D_compensation}
\end{figure}

Micromotion can be minimized in two dimensions using a single probe laser beam by using the resolved sideband method \cite{Berkeland1998} together with interferometry method~A.
This is shown in Fig.~\ref{fig_1_beam_2D_compensation}.
In Fig.~\ref{fig_1_beam_2D_compensation}(a) the interferometry method is conducted using a horizontal laser beam, and $\phi_\mathrm{PD}$ is sensitive to the horizontal component of $\vec{E}$, which is varied by changing the voltage applied to the ``horizontal'' compensation electrode.
In Fig.~\ref{fig_1_beam_2D_compensation}(c) the resolved sideband method is conducted using the same horizontal laser beam, and the sideband amplitude is sensitive to the vertical component of $\vec{E}$, which is varied by changing the voltage applied to the ``vertical'' compensation electrode.
Similar results are observed when using a vertical laser beam in Fig.~\ref{fig_1_beam_2D_compensation}(b) and (d).

\begin{figure}[ht!]
\centering
\includegraphics[width=0.8\columnwidth]{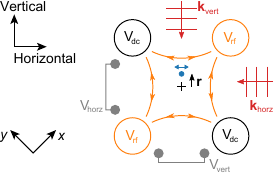}
\caption{Schematic of the experimental setup, showing a slice through the linear Paul trap in the plane of the oscillating electric field.
The trap's oscillating electric quadrupole field (orange lines) is produced by applying voltages to four electrodes (large circles).
An unwanted dipolar field along the vertical direction displaces the ion equilibrium position vertically, with displacement $\vec{r}$ from the trap centre (black cross).
This can be detected using interferometry method~A with the vertical laser beam, or using the resolved sideband method with the horizontal laser beam, since at the new position the ion (blue dot) experiences a horizontal oscillating dipole field which drives horizontal micromotion (blue arrow).
Each compensation electrode consists of a pair of rods (grey circles).
The ion's secular motion eigenmodes are orientated along $x$ and $y$.
}
\label{fig_sb_method}
\end{figure}

These results can be understood with the aid of Fig.~\ref{fig_sb_method}.
Using a horizontal laser beam and interferometry method~A, the results are sensitive to the horizontal component of $\vec{E}$, which displaces the ion equilibrium position horizontally.
Using a horizontal laser beam and the resolved sideband method, the results are sensitive to the vertical component of $\vec{E}$, which displaces the ion equilibrium position vertically, at the new equilibrium position the ion experiences a horizontal oscillating dipole field, which drives horizontal micromotion.

\subsection{Applying the interferometry methods in linear Paul traps with non-degenerate radial frequencies}
Micromotion minimization techniques that involve monitoring an ion's position when the trap stiffness is changed become more sensitive when larger changes of the trap stiffness are used.
However, in a linear Paul trap, if the trap stiffness is reduced to the point where the ion is barely trapped, and if non-degenerate radial secular frequencies are used, these techniques risk becoming overwhelmingly sensitive to the offset field $\vec{E}$ along just one direction.

\begin{figure}[ht!]
\centering
\includegraphics[width=\columnwidth]{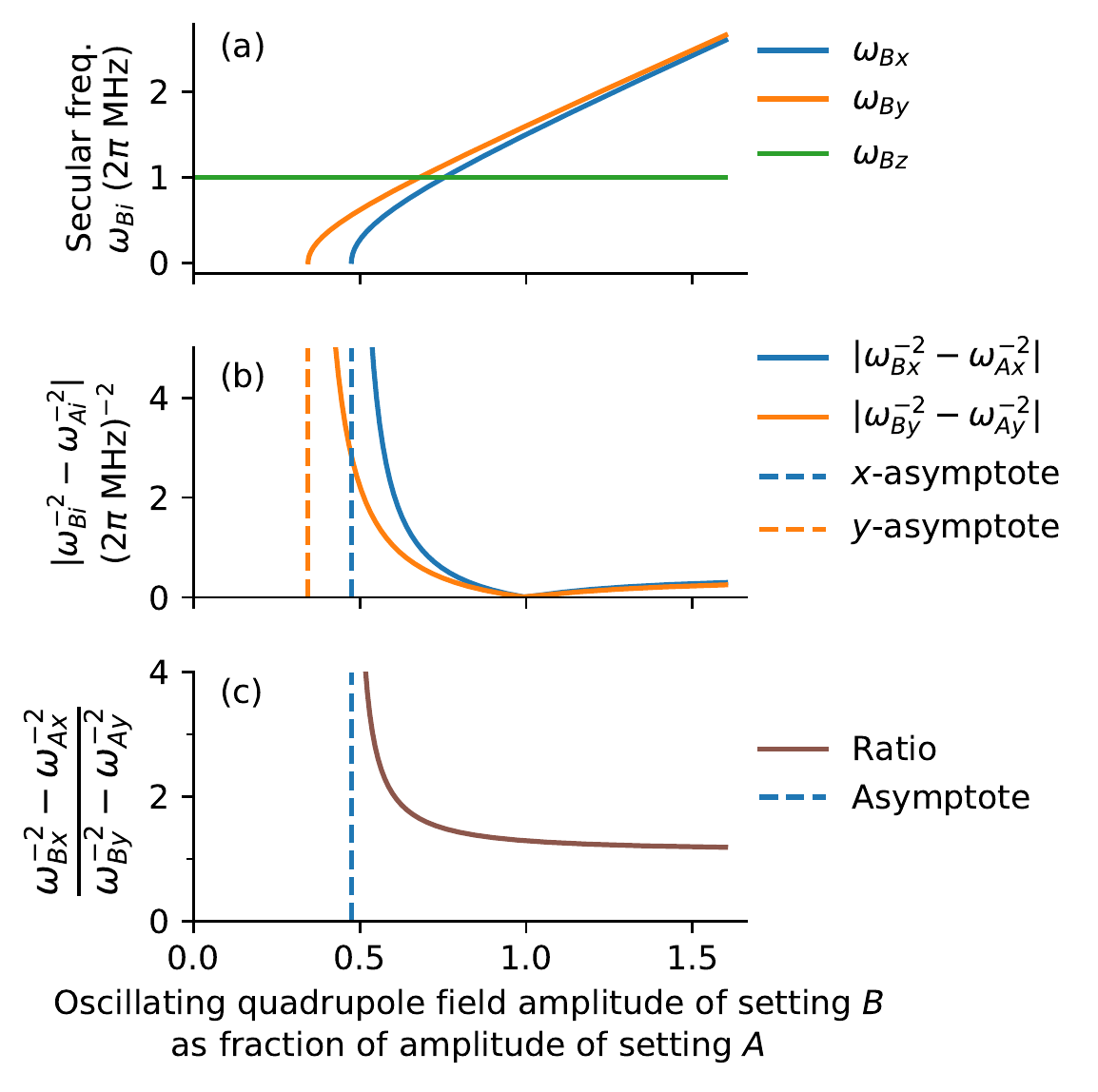}
\caption{When the radial trapping frequencies in a linear Paul trap are non-degenerate with $\omega_x < \omega_y$, an offset field along the $x$-direction causes a larger change of ion equilibrium position than an offset field of the same magnitude along the $y$-direction.
This difference diverges as the amplitude of the oscillating quadrupole field is reduced.
In the figure the trap stiffness is changed between initial settings $A$ with $\{\omega_x, \omega_y, \omega_z\}/2\pi = \{1.5, 1.6, 1.0\}\,\mathrm{MHz}$ and settings $B$ by changing the amplitude of the oscillating quadrupole field.
(a)~As the oscillating quadrupole field amplitude during setting $B$ is decreased, $\omega_{Bx}\rightarrow0$ before $\omega_{By}\rightarrow0$.
(b)~The ion displacement ${r}_{ABi}$ due to an offset field $E_i$ depends on $\omega_{Bi}^{-2}-\omega_{Ai}^{-2}$ [see Eq.~(\ref{eq_del_u})].
This quantity diverges as $\omega_{Bi}\rightarrow0$.
(c)~The ratio of $\omega_{Bx}^{-2}-\omega_{Ax}^{-2}$ to $\omega_{By}^{-2}-\omega_{Ay}^{-2}$, indicating the relative displacements caused by an offset field, diverges as the trap stiffness during setting $B$ is reduced and $\omega_{Bx}\rightarrow0$.
}
\label{fig_non_degeneracy}
\end{figure}

We illustrate this behaviour in Fig.~\ref{fig_non_degeneracy}.
The calculations show how the trap stiffnesses respond when the amplitude of the linear Paul trap's oscillating quadrupole field is changed, from an initial setting $A$, with non-degenerate trap stiffnesses $\{\omega_{Ax},\omega_{Ay}\}/2\pi=\{1.5,1.6\}\,\mathrm{MHz}$ along the radial directions and $\omega_{Az}/2\pi=1.0\,\mathrm{MHz}$ along the axial direction, to a trap setting $B$.
As we decrease the amplitude of the trap's oscillating quadrupole field $\omega_{Bx}\rightarrow0$ before $\omega_{By}\rightarrow0$ and thus the quantity $\omega_{Bx}^{-2}-\omega_{Ax}^{-2}$ diverges before $\omega_{By}^{-2}-\omega_{Ay}^{-2}$ diverges.
These quantities describes the response of $\vec{r}_{AB}$ to $\vec{E}$ [Eq.~(\ref{eq_del_u})] and impact the direction $\vec{d}$ along which the interferometry method is sensitive to $\vec{E}$ [Eqs.~(\ref{eq_d}) and (\ref{eq_d2}), and in Appendix~\ref{appendix_method_c} Eqs.~(\ref{eq_d3}) and (\ref{eq_d4})].
As a result, when Method~A is used with a beam that has a projection onto both the $x$ and $y$ axes, and when the oscillating quadrupole field amplitude is reduced to the point where the ion is barely trapped ($\omega_{Bx}\approx 0$) the technique effectively becomes sensitive to only $E_x$.
A much higher sensitivity to $E_x$ than to $E_y$ also appears if the radial stiffnesses are reduced by increasing the amplitude of the static quadrupole field which provides axial confinement.

\begin{figure}[ht!]
\centering
\includegraphics[width=\columnwidth]{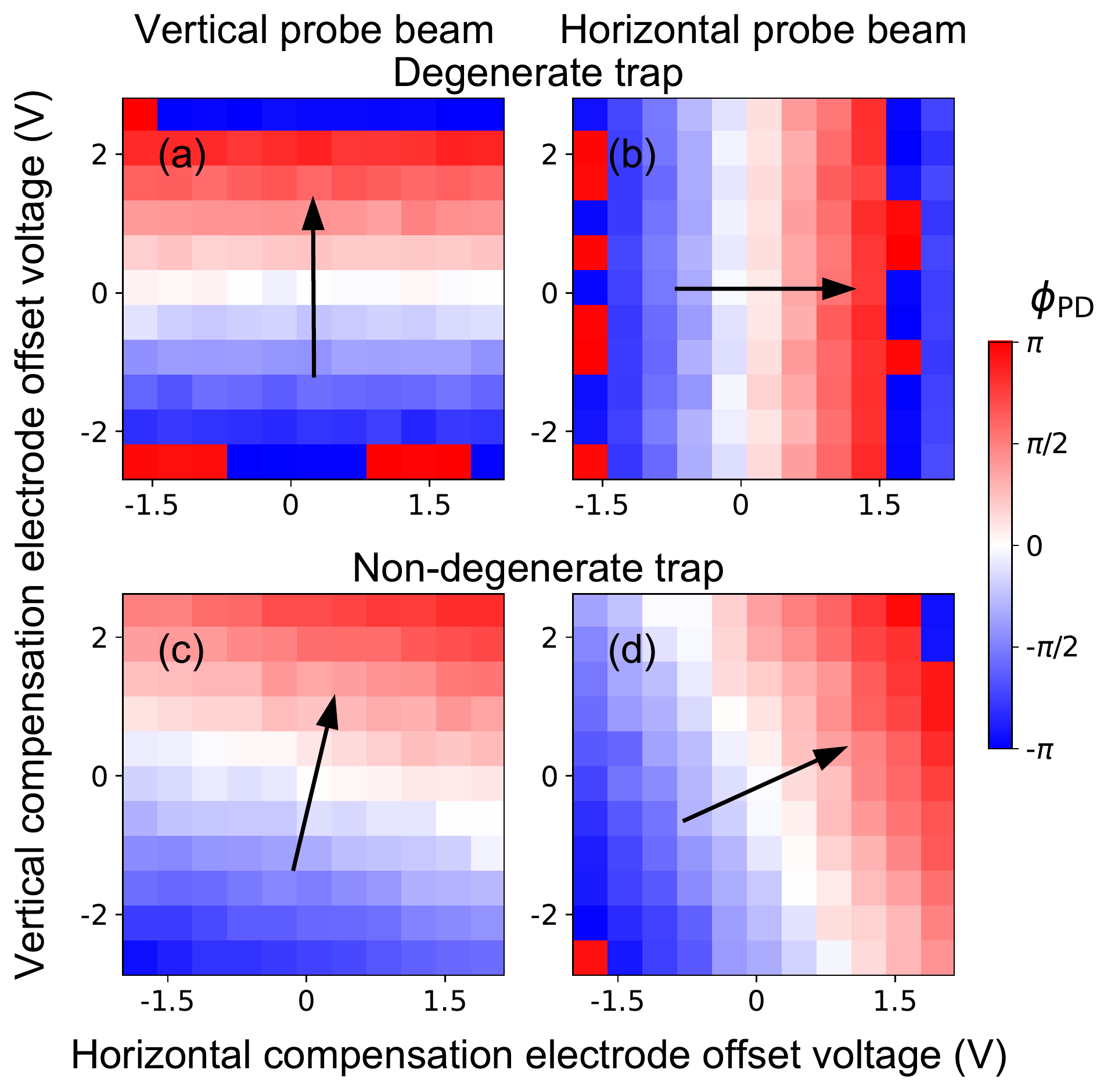}
\caption{2D micromotion compensation with degenerate secular frequencies compared with the case of non-degenerate secular frequencies.
In (a) [(b)] the secular frequencies are degenerate, and $\phi_\mathrm{PD}$ is sensitive to the vertical [horizontal] component of $\vec{E}$ along $\tfrac{1}{\sqrt{2}}(\hat{x}+\hat{y})$ [$\tfrac{1}{\sqrt{2}}(\hat{x}-\hat{y})$] when measured using a vertical [horizontal] beam.
In (c) and (d) the secular frequencies are non-degenerate, with $\omega_x < \omega_y$, and as a result the measurements of $\phi_\mathrm{PD}$ become more sensitive to $\vec{E}$ along the $x$-direction, i.e.\ the ``horizontal~$+$~vertical'' direction, as described by Eq.~(\ref{eq_d}) and Fig.~\ref{fig_non_degeneracy}.
}
\label{fig_seq1_2D_comparison}
\end{figure}

We illustrate this sensitivity difference by conducting experiments using Method~A as $E_x$ and $E_y$ are changed, using two different laser beams which each have projections onto the $x$ and $y$ axes. The beam directions are shown in the schematic in Fig.~\ref{fig_sb_method}.
The results are shown in Fig.~\ref{fig_seq1_2D_comparison}.
In Figs.~\ref{fig_seq1_2D_comparison}(a) and (b) the secular frequencies are degenerate ($\omega_x=\omega_y$), and $\phi_\mathrm{PD}$ depends on the vertical (horizontal) component of $\vec{E}$ when a vertical (horizontal) probe beam is used; the orthogonal beams are sensitive to orthogonal components of $\vec{E}$.
In Figs.~\ref{fig_seq1_2D_comparison}(c) and (d) the secular frequencies are non-degenerate with $\omega_x < \omega_y$ and the method becomes more sensitive to $E_x$ than to $E_y$.
As a result, the orthogonal beams are sensitive to non-orthogonal components of $\vec{E}$.

If a higher sensitivity to $E_x$ than to $E_y$ is problematic, Method~C (Appendix~\ref{appendix_method_c}) may be useful; it allows the direction of sensitivity $\vec{d}$ to be tuned.
Another solution is to implement Method~A using a probe beam propagating along the $y$-axis, with no projection onto the $x$-axis.
However, in most setups the electrode geometry obstructs optical access along the directions of secular motion.
A third solution is to calculate superpositions of the phases measured via Method~A using probe beams from different directions, for instance a weighted sum (difference) of the phases measured with the horizontal and vertical beams is sensitive to $E_x$ ($E_y$).
Alternatively one can use Method~B, with two beams whose wavevector difference $\vec{k}_\alpha-\vec{k}_\beta$ has no $x$-component [see Eq.~(\ref{eq_d2})].

\subsection{Minimization of axial micromotion in a linear Paul trap}

In an ideal linear Paul trap there is no RF electric field along the trap symmetry axis ($z$ direction) $\tilde{E}_z=0$.
In physical linear Paul traps, $\tilde{E}_z$ is non-zero because of the finite size of the trap electrodes, among other reasons \cite{Herschbach2012, Pyka2014, Keller2015, Keller2019}.
A non-zero $\tilde{E}_z$ drives ion micromotion along $z$.
Usually $\tilde{E}_z$ vanishes only at a single point, and with increasing distance from this point along $z$, $|\tilde{E}_z|$ increases \cite{Pyka2014} and the extent of axial micromotion increases.
Thus, the null point can be found using, for example, the resolved sideband method \cite{Berkeland1998} with a laser beam propagating along the $z$ direction.

\begin{figure}[ht!]
\centering
\includegraphics[width=0.9\columnwidth]{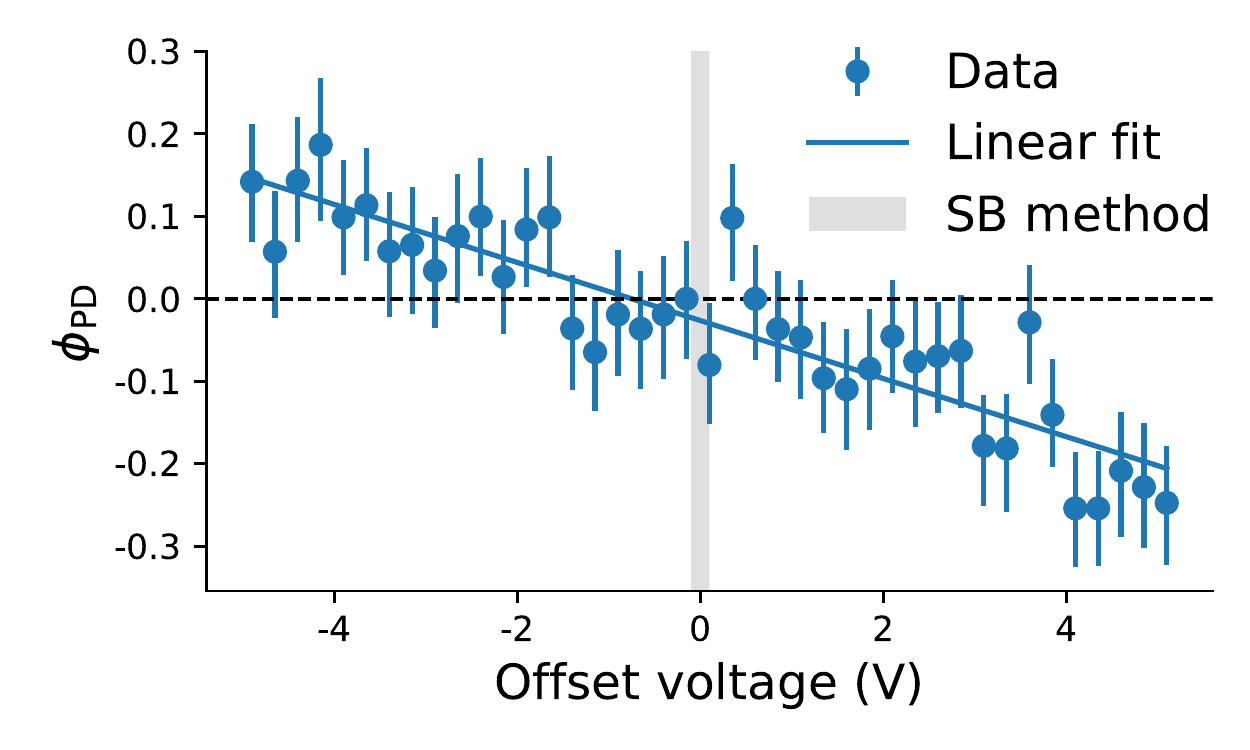}
\caption{The interferometry sequences enable axial micromotion compensation in a linear Paul trap.
The axial component of $\vec{E}$ is varied by offsetting the voltage applied to an endcap electrode, and $\phi_\mathrm{PD}$ responds linearly.
$\phi_\mathrm{PD}$ is measured using Method~A with a beam propagating along the axial direction.
Error bars represent quantum projection noise (1$\sigma$ confidence interval).
The shaded area indicates the $1\sigma$ uncertainty in the estimate obtained using the resolved sideband method.
}
\label{fig_axial_MM}
\end{figure}

Because the extent of axial micromotion and thus the kinetic energy associated with it increase with distance along $z$ from the null, $\tilde{E}_z$ introduces a trapping pseudopotential along $z$.
This pseudopotential contributes to the axial confinement, and this means that axial micromotion can be minimised using methods which are sensitive to the change of ion equilibrium position along $z$ when $\omega_z$ is changed \cite{Gloger2015}.
And so, we demonstrated that interferometry method~A can be used to minimize axial micromotion:

We varied the $z$-component of $\vec{E}$ (and thus we varied $\tilde{E}_z$) by changing the voltage applied to an endcap electrode, and we measured the linear response of $\phi_\mathrm{PD}$ using a laser beam with wavevector $\vec{k}$ largely along the $z$-direction (it propagates through holes in the endcap electrodes).
We changed $\omega_z$ during the pulse sequence by changing the amplitude of the oscillating electric quadrupole field; this can also be achieved by changing the amplitude of the static electric quadrupole field.
The results are shown in Fig.~\ref{fig_axial_MM}.
The zero-offset voltage was determined using the resolved sideband method \cite{Berkeland1998}; the optimal voltage determined using the interferometry method and the optimal voltage determined using the resolved sideband method do not perfectly agree.
This mismatch may have resulted from a small projection of the probing laser beam onto the $x$ and $y$ directions (the plane of the oscillating quadrupole field) together with non-zero $x$- and $y$-components of $\vec{E}$.

\section{Demonstration of quantum clock synchronization protocols} \label{sec_ticking_qubit}
Method~A has much in common with two quantum clock synchronization protocols \cite{Chuang2000, deBurgh2005}.
Synchronizing distant clocks is important for engineering and metrology.
It is also of fundamental interest in physics, falling within the field of reference frame alignment \cite{Bartlett2007}.
Suppose Alice and Bob want to synchronize their clocks, which are known to tick at the same rate:
Eddington's protocol \cite{Eddington1924} involves Alice synchronising a watch to her clock, and then mailing the watch to Bob, who synchronises his own clock to the watch.
Chuang \cite{Chuang2000} proposed a quantum version of Eddington's protocol, in which Alice sends a quantum watch to Bob, namely a ticking qubit.
In this protocol Alice and Bob each apply a $\pi/2$ pulse on the qubit before the state of the qubit is measured.
Importantly, the phase of each $\pi/2$ pulse is relative to Alice's and Bob's clocks respectively.

\begin{figure}[ht!]
\centering
\includegraphics[width=\columnwidth]{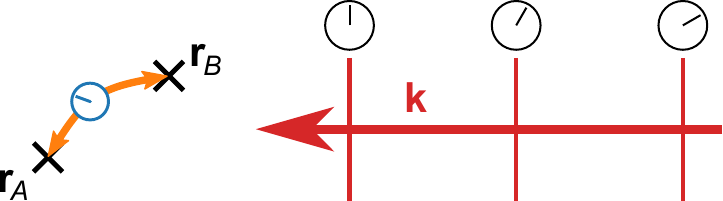}
\caption{Method~A is related to quantum versions of Eddington's clock synchronization protocol \cite{Chuang2000,deBurgh2005}:
Alice and Bob (phantoms at $\vec{r}_A$ and $\vec{r}_B$) have an unknown phase difference $\phi_{\mathrm{PD}}$ between their clocks (the oscillating laser field at their positions).
They exchange a ticking qubit (a trapped ion), and they each perform rotations on it.
By measuring the difference between their rotation axes they learn $\phi_{\mathrm{PD}}$.
}
\label{fig_clock_sync}
\end{figure}

The sequence of Method~A with $M=1$ is equivalent to Chuang's protocol.
In this sequence the trapped ion equilibrium position changes from $\vec{r}_A$ (Alice's location) to $\vec{r}_B$ (Bob's location) when the trap stiffness is changed.
We identify the optical field at $\vec{r}_A$ as Alice's clock, and the optical field at $\vec{r}_B$ as Bob's clock (these clocks tick incredibly fast, at over 400\,THz).
The asynchronicity of the clocks is due to the phase difference $\phi_\mathrm{PD}$ between the optical field at $\vec{r}_A$ and the optical field at $\vec{r}_B$ [see Eq.~(\ref{eq_phi_T_k_r_AB})].
During the sequence we first apply a $\pi/2$ pulses on a ticking ion qubit at position $\vec{r}_A$ (the pulse phase is determined by Alice's clock) then we move the qubit to $\vec{r}_B$ and apply another $\pi/2$ pulse (the pulse phase is determined by Bob's clock), before measuring the state of the qubit.
Measurements allow us to calculate $\phi_\mathrm{PD}$ and thus ``synchronise the clocks'', as shown in Fig.~\ref{fig_seq1}.
In Fig.~\ref{fig_clock_sync} we illustrate the relationship between Method A and Chuang's protocol.

De~Burgh and Bartlett \cite{deBurgh2005} improved on Chuang's protocol.
They proposed that Alice and Bob perform multiple exchanges of the qubit, and apply multiple pulses on the qubit, to more accurately determine $\phi_\mathrm{PD}$.
Method~A with $M>1$ is equivalent to this protocol, and the data in Figs.~\ref{fig_seq_multi_pulse_data} and \ref{fig_rpe} demonstrates the enhancement gained from this protocol over the two-pulse protocol.\footnote{Chuang's paper \cite{Chuang2000} includes a protocol with a sub-SQL scaling, however, this protocol requires a set of ticking qubits, whose frequencies span an exponentially-large range.}

Within the framework developed in this section, we can describe Method~B as a protocol to synchronise two oscillators (i.e.\ two laser fields) which are at the same position, using a ticking qubit.

\section{Conclusion}
We introduce and demonstrate interferometry pulse sequences for minimizing the magnitude of a stray electric field $\vec{E}$ in a trapped ion experiment.
These sequences allow $|\vec{E}|$ to be minimized to state-of-the-art levels quickly, with modest experimental requirements.
These methods will be particularly useful in trapped ion precision spectroscopy experiments \cite{Keller2015, Brewer2019}, hybrids systems of neutral atoms and trapped ions \cite{Grier2009,Schmid2010,Zipkes2010,Feldker2020}, and experiments using highly-polarizable Rydberg ions \cite{Higgins2019, Feldker2015}, which are very sensitive to effects caused by stray fields.

We demonstrate that quantum phase estimation techniques can be used to minimize $|\vec{E}|$ with a scaling below the standard quantum limit.
This constitutes a real-world case in which quantum metrology provides a significant enhancement.
We also show that the results can be robust against laser detuning and pulse area errors.

By using one of the sequences presented here together with the resolved sideband method we minimize $|\vec{E}|$ in 2D using a single probe beam.
This approach will be useful in experiments with restricted optical access, such as cavity QED experiments \cite{Sterk2012, Steiner2013, Stute2013} and surface trap experiments \cite{Brown2011, Harlander2011, Wilson2014, Kumph2016, Mehta2020, Niffenegger2020}.

We reduced $|\vec{E}|$ beyond state-of-the-art levels quickly.
$|\vec{E}|$ could be reduced much further and much more quickly in a setup with a longer coherence time (allowing longer sequences) and with finer control of the trap stiffness (allowing larger stiffness changes).

In trapped ion precision spectroscopy experiments usually just a single ion is probed.
Scaling up precision spectroscopy experiments to many ions enables faster interrogation \cite{Champenois2010, Pyka2014, Arnold2015, Keller2019}.
In a  many-ion system the offset field $\vec{E}$ would ideally be measured and countered for each of the ions.
The methods presented here will work in a system of many ions, provided that the ions do not unexpectedly switch positions during the sequences.
Further, by probing a system of entangled ions, it might be possible to precisely measure offset fields even faster \cite{Gilmore2021}.

The methods we introduce can also be used when the states which get excited are separated by a Raman transition or a multi-photon transition.
To achieve the highest sensitivity the laser beams should be orientated to give the largest effective wavevector.

The dominant cause of excess micromotion is usually a slowly-varying dipole field $\vec{E}$ at the null of the oscillating quadrupole field.
However, excess micromotion can also arise when the oscillating voltages applied to the trap electrodes are out of phase, this is called quadrature micromotion.
Measurements sensitive to $\vec{r}_{AB}$, such as the techniques presented here, do not give information about quadrature micromotion.
Quadrature micromotion can instead be characterised using other methods \cite{Berkeland1998, Keller2015} and it can be avoided by careful trap design and fabrication \cite{Herschbach2012,Pyka2014,Chen2017}.

Finally, our work demonstrates quantum versions of Eddington's clock synchronisation protocol \cite{Chuang2000, deBurgh2005}, linking trapped ion experiments to the problem of reference frame alignment \cite{Bartlett2007}.

\section*{Acknowledgements}
We thank Holger Motzkau for designing and making the bias tee in Fig.~\ref{fig_electronics}.
We thank Ferdinand Schmidt-Kaler for making us aware of ref.~\cite{Kotler2011}.
This work was supported by the Knut \& Alice Wallenberg Foundation (Photonic Quantum Information and through the Wallenberg Centre for Quantum Technology [WACQT]), the QuantERA ERA-NET Cofund in Quantum Technologies (ERyQSenS), and the Swedish Research Council (Trapped Rydberg Ion Quantum Simulator and grant number 2020-00381).

\section*{Author contributions}
GH developed the methods, planned and conducted the experiments, analysed the data and wrote the manuscript.
All authors contributed to the experimental setup, discussed the results and gave feedback on the manuscript.

\appendix

\section{Statistical uncertainty in \texorpdfstring{$\phi_\mathrm{T}$}{phase} estimates} \label{appendix_a}
Eqs.~(\ref{eq_phiT_2}) and (\ref{eq_phiT_3}) describe ways to estimate $\phi_\mathrm{T}$ using the arcsin and arctan2 functions.
In the following sections we illustrate why the statistical error in the $\phi_\mathrm{T}$ estimate using the arctan2 function depends on $\phi_\mathrm{T}$.
Then we compare the statistical errors of the $\phi_\mathrm{T}$ estimates using the arcsin and arctan2 functions.

\subsection{Variation of the \texorpdfstring{$\phi_\mathrm{T}$}{phase} estimate's statistical uncertainty with the value of \texorpdfstring{$\phi_\mathrm{T}$}{the phase}} \label{appendix_statistical_uncertainty}
\begin{figure}[ht!]
\centering
\includegraphics[width=\columnwidth]{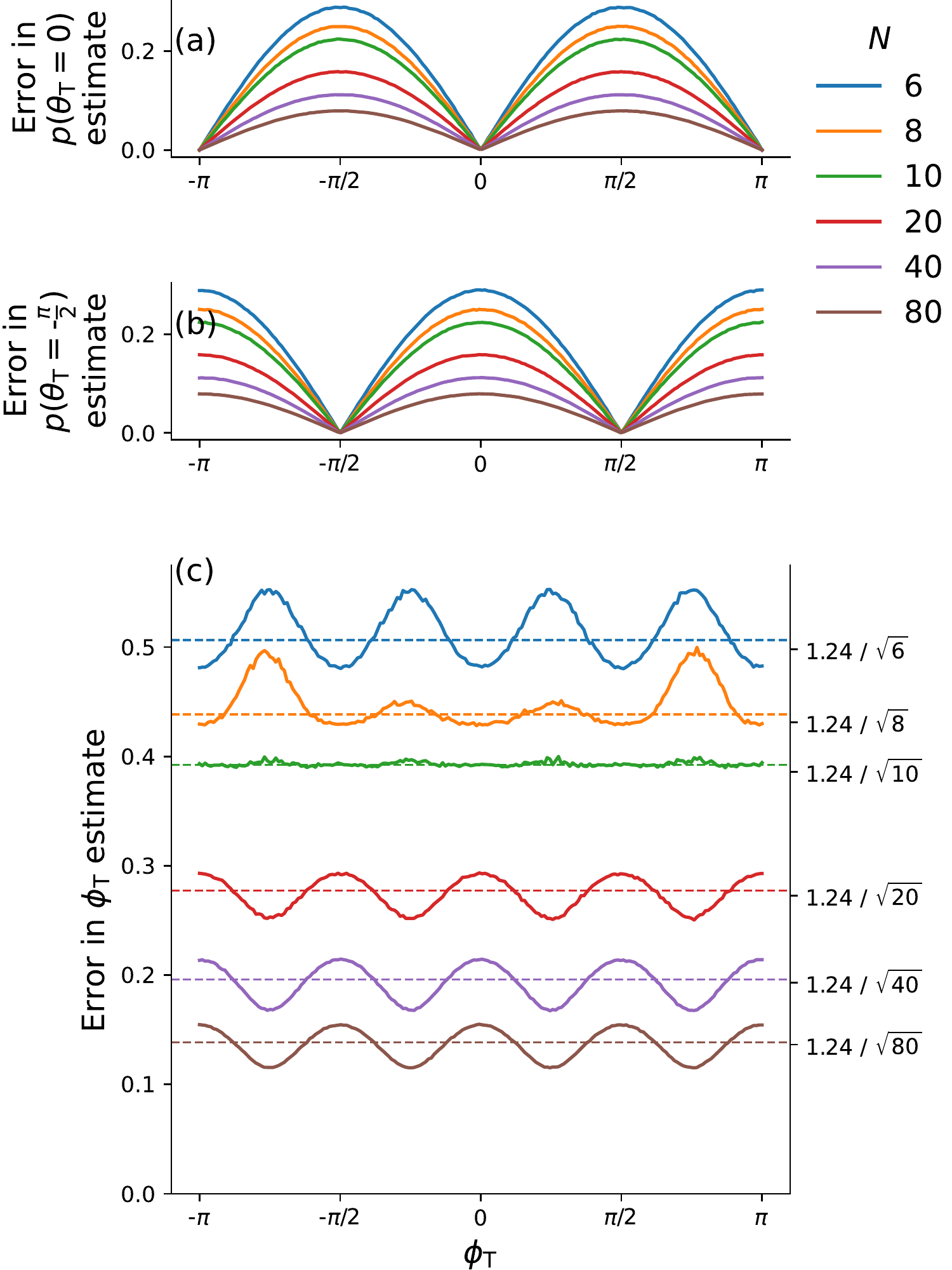}
\caption{Probing the statistical error of an estimate of $\phi_\mathrm{T}$ using numerical simulations.
(a)~and (b)~show how the statistical uncertainties of measurements of the probability $p$ depend on $\phi_\mathrm{T}$.
(a)~and (b)~use $\theta_\mathrm{T}=0$ and $-\pi/2$ respectively.
$p$ depends on $\phi_\mathrm{T}$ according to Eq.~(\ref{eq_pe_cos_theta_T_phi_T}).
The statistical uncertainty of a $p$ measurement is minimal when the true value is 0 or 1 ($\phi_\mathrm{PD}+\phi_\mathrm{c}=0$ or $\pm \pi$) and the uncertainty is maximal when the true value is 0.5 ($\phi_\mathrm{PD}+\phi_\mathrm{c}=\pm \pi/2$).
The uncertainties decrease as the number of samples used to generate the estimates is increased.
In (a) and (b) $N/2$ samples were used to calculate the results.
In (c) estimates of $\phi_\mathrm{T}$ were constructed from $N/2$ simulated measurements of $p(\theta_\mathrm{T}=0)$ and $N/2$ simulated measurements of $p(\theta_\mathrm{T}=-\tfrac{\pi}{2})$ using Eq.~(\ref{eq_phiT_3}).
The statistical error of the $\phi_\mathrm{T}$ estimates depends on the true value of $\phi_\mathrm{T}$, and the average error is $\approx 1.24/\sqrt{N}$.
}
\label{fig_statistical_uncertainty}
\end{figure}

The statistical uncertainty in an estimate of $\phi_\mathrm{T}$ using Eq.~(\ref{eq_phiT_3}) is $\approx \frac{1.24}{\sqrt{N}}$.
The uncertainty varies depending on the true value of $\phi_\mathrm{T}$, as shown by the simulated data in Fig.~\ref{fig_statistical_uncertainty}.

To obtain this data, we simulated binomial trials with success probabilities given by Eq.~(\ref{eq_pe_cos_theta_T_phi_T}), $N/2$ trials using $\theta_\mathrm{T}=0$ and $N/2$ trials using $\theta_\mathrm{T}=-\tfrac{\pi}{2}$.
These trials gave estimates of $p(\theta_\mathrm{T}=0)$ and $p(\theta_\mathrm{T}=-\tfrac{\pi}{2})$.
From these estimates we estimated $\phi_\mathrm{T}$ using Eq.~(\ref{eq_phiT_3}).
We repeated this several thousand times for each $\phi_\mathrm{T}$ value, and using different values of $N$ between 6 and 80.
The root-mean-square error in the estimates of $p(\theta_\mathrm{T}=0)$, $p(\theta_\mathrm{T}=-\tfrac{\pi}{2})$ and $\phi_\mathrm{T}$ are shown in Fig.~\ref{fig_statistical_uncertainty}(a), (b) and (c) respectively.

The statistical error in the estimates of $p$ are minimal when the true $p$ values are either 0 or 1 (in the same way that a coin toss that is certain to return heads will \textit{always} return heads).
Because the statistical errors of the $p$ estimates depend on $\phi_\mathrm{T}$, this gets carried over so that the statistical error of the $\phi_\mathrm{T}$ estimate depends on the true value of $\phi_\mathrm{T}$.

We note a change of behaviour around $N=10$: when $N>10$ the statistical uncertainty in the estimate of $\phi_\mathrm{T}$ using Eq.~(\ref{eq_phiT_3}) is highest when the true value of $\phi_\mathrm{T}$ is $0$, whereas when $N<10$ the uncertainty in the estimate is lowest when the true value is $0$.
We expect this change of behaviour arises as follows: The statistical uncertainties in estimates of probabilities $p(\theta_T=0)$ and $p(\theta_T=-\tfrac{\pi}{2})$ grow as $N$ is decreased (as shown in Fig.~\ref{fig_statistical_uncertainty}). Because a $\phi_\mathrm{T}$ estimate is produced from the estimates of $p(\theta_T=0)$ and $p(\theta_T=-\tfrac{\pi}{2})$ using a nonlinear function [Eq.~(\ref{eq_phiT_3})], the statistical uncertainty in a $\phi_\mathrm{T}$ estimate has a nonlinear dependence on the statistical uncertainties in the estimates of $p(\theta_T=0)$ and $p(\theta_T=-\tfrac{\pi}{2})$, and thus on $N$.
We expect the change of behaviour of the uncertainty in $\phi_\mathrm{T}$ estimates at $N=10$ arises from this nonlinear dependence.

\subsection{Comparing the statistical errors in \texorpdfstring{$\phi_\mathrm{PD}$}{phase} estimates obtained using arcsin and arctan2 functions} \label{appendix_arcsin}
\begin{figure}[ht!]
\centering
\includegraphics[width=\columnwidth]{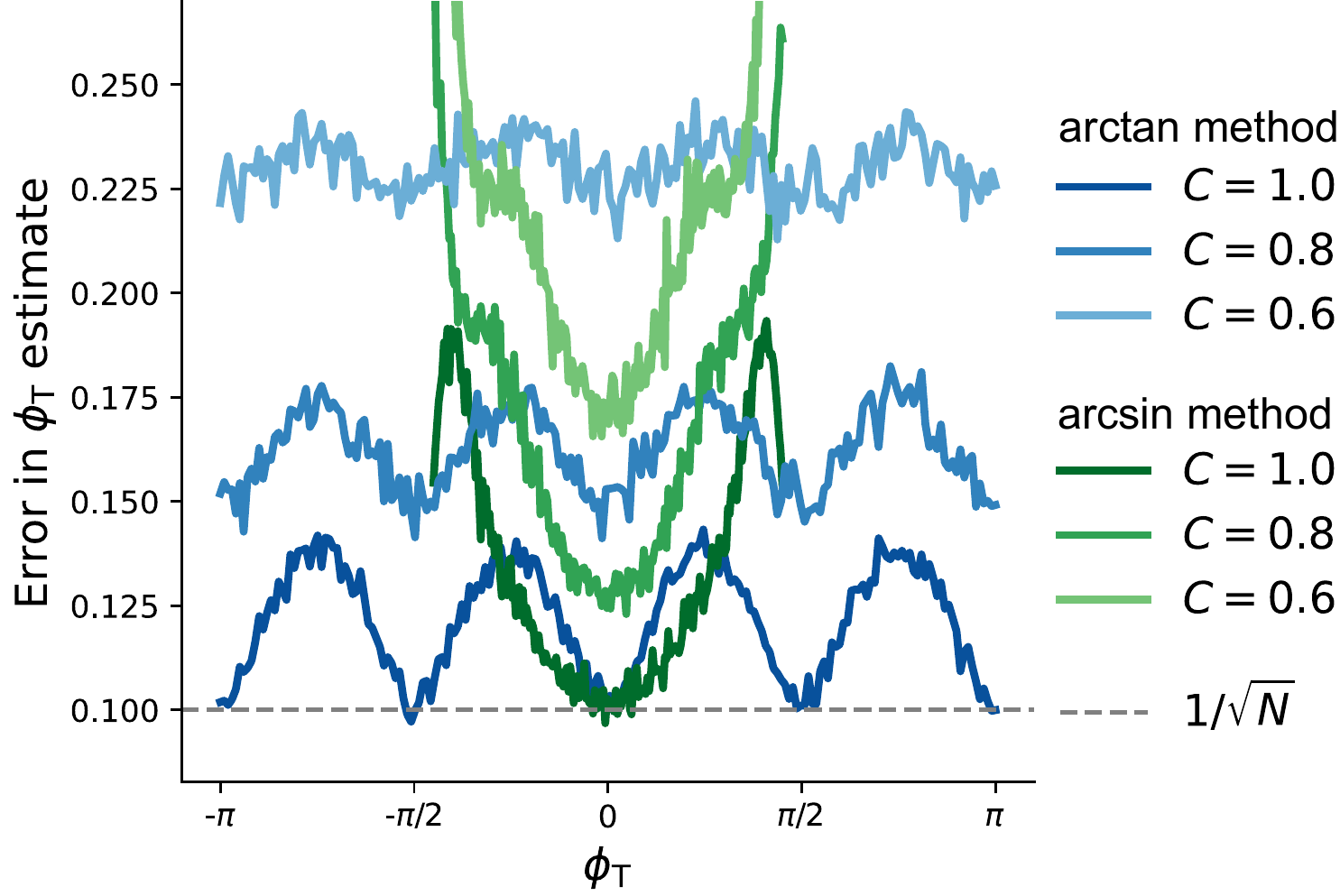}
\caption{The statistical uncertainty in the estimates of $\phi_\mathrm{T}$ increases as the contrast $C$ decreases.
Estimates obtained using the arcsin method are more precise than estimates obtained using the arctan2 method when the contrast $C$ is lower and when $\phi_\mathrm{T}$ is small.
The $\phi_\mathrm{T}$ estimates were constructed using $N=100$ simulated samples.
The dashed grey line indicates the standard quantum limit \cite{Giovannetti2004}.
}
\label{fig_statistical_uncertainty_arcsin}
\end{figure}

When $\phi_\mathrm{T}$ is small, estimates of $\phi_\mathrm{T}$ obtained using the arcsin function have lower statistical errors than estimates of $\phi_\mathrm{T}$ obtained using the arctan2 function.
This is especially true when the contrast $\mathcal{C}$ of the oscillation described by Eq.~(\ref{eq_pe_cos_theta_T_phi_T}) is reduced.
We show this using the simulated data in Fig~\ref{fig_statistical_uncertainty_arcsin}.
The data in the figure was obtained by simulations, in the same way as in the previous section.
The disadvantage of the arcsin approach is that it can only return an estimate of $\phi_\mathrm{PD}$ within the range [$-\pi/2, \pi/2$], while the arctan2 approach can return an estimate within the range [$-\pi, \pi$].

Using the arctan approach described by Eq.~(\ref{eq_phiT_3}) the statistical uncertainty is maximal at $\phi_\mathrm{T}=0$ (for $N>10$), as shown in Fig.~\ref{fig_statistical_uncertainty}.
For a better comparison between the arcsin and arctan2 approaches, the data in Fig.~\ref{fig_statistical_uncertainty_arcsin} uses the estimate
\begin{equation} \label{eq_phiT_4}
\phi_\mathrm{T} = \mathrm{arctan2}\left[p(\theta_T=\tfrac{\pi}{4})-\tfrac{1}{2}, p(\theta_T=\tfrac{3\pi}{4})-\tfrac{1}{2}\right] - \tfrac{3\pi}{4}
\end{equation}
which has a minimum statistical uncertainty at $\phi_\mathrm{T}=0$.
If $\phi_\mathrm{PD}$ is known to be small Eq.~(\ref{eq_phiT_4}) will give a better estimate than Eq.~(\ref{eq_phiT_3}).

\section{Method C -- Sequence using trap stiffness changes and multiple laser beams} \label{appendix_method_c}
Here we introduce a sequence which allows the direction $\vec{d}$, along which the measurement results are sensitive to the offset field $\vec{E}$, to be tuned.
The sequence comprises four subsets of pulses, driven by two different laser beams and using two different trap stiffness settings.
The sequence consists of two $\pi/2$ pulses separated by $(M-1)$ $\pi$ pulses, just as the sequences of Methods A and B.

Pulses within the four subsets are driven by the laser fields with wavevectors $\vec{k}_\alpha$, $\vec{k}_\alpha$, $\vec{k}_\beta$ and $\vec{k}_\beta$ respectively, when trap stiffnesses $A$, $B$, $A$ and $B$ are used.
The first and second subsets each have total pulse area $M_\alpha \pi/2$, while the third and fourth subsets each have total pulse area $M_\beta \pi/2$, where $M_\alpha$ and $M_\beta$ are integers which satisfy $M=M_\alpha + M_\beta$.
Pulses from the first and third subsets have odd indices $j$, while pulses from the second and fourth subsets have even indices $j$.
From Eqs.~(\ref{eq_phi_T}) and (\ref{eq_Phi_alpha_A_Phi_alpha_B})
\begin{align}
\begin{split}
\phi_\mathrm{T} &= M_\alpha \left( \Phi_{\alpha A} - \Phi_{\alpha B} \right) +M_\beta \left( \Phi_{\beta A} - \Phi_{\beta B} \right) \\
&= \left( M_\alpha \vec{k}_\alpha + M_\beta \vec{k}_\beta \right) \cdot \left( \vec{r}_A - \vec{r}_B \right)
\end{split}
\end{align}
which is sensitive to $\vec{E}$ along $\vec{d}$, which has components
\begin{equation} \label{eq_d3}
d_i = \left( M_\alpha k_{\alpha i} + M_\beta k_{\beta i} \right) \left( \frac{1}{{\omega_{Ai}}^2} - \frac{1}{{\omega_{Bi}}^2} \right)
\end{equation}
By changing $M_\alpha$ and $M_\beta$ the direction $\vec{d}$ in which $\phi_\mathrm{T}$ is sensitive to $\vec{E}$ can be adjusted.
In this sequence the trap stiffness is alternated between stiffnesses $A$ and $B$ between each pulse; there are $M_\alpha+M_\beta=M$ changes of the trap stiffness during the sequence.

If instead the pulses of the third subset have even indices $j$ while the pulses of the fourth subset have odd indices $j$ then
\begin{equation} \label{eq_phi_T_general_method}
\phi_\mathrm{T} = M_\alpha \left( \Phi_{\alpha, A} - \Phi_{\alpha, B} \right) -M_\beta \left( \Phi_{\beta, A} - \Phi_{\beta, B} \right)
\end{equation}
and the direction $\vec{d}$ becomes
\begin{equation} \label{eq_d4}
d_i = \left( M_\alpha k_{\alpha i} - M_\beta k_{\beta i} \right) \left( \frac{1}{{\omega_{Ai}}^2} - \frac{1}{{\omega_{Bi}}^2} \right)
\end{equation}
and the number of trap stiffness changes can be reduced to $|M_\alpha - M_\beta|$ if $\left( M_\alpha+M_\beta \right)$ is odd and $\text{max}\{|M_\alpha - M_\beta|,2\}$ if $\left( M_\alpha+M_\beta \right)$ is even.
Note that when $M_\alpha=M_\beta$ the phase in Eq.~(\ref{eq_phi_T_general_method}) is the same as the phase determined using Method~B.

Adding a fifth and sixth subset of pulses driven by a third laser beam with a wavevector $\vec{k}_\gamma$ may allow $\vec{d}$ to be varied in three dimensions.

\section{Control of the trap stiffness} \label{sec_electronics}
\begin{figure}[ht]
\centering
\includegraphics[width=\columnwidth]{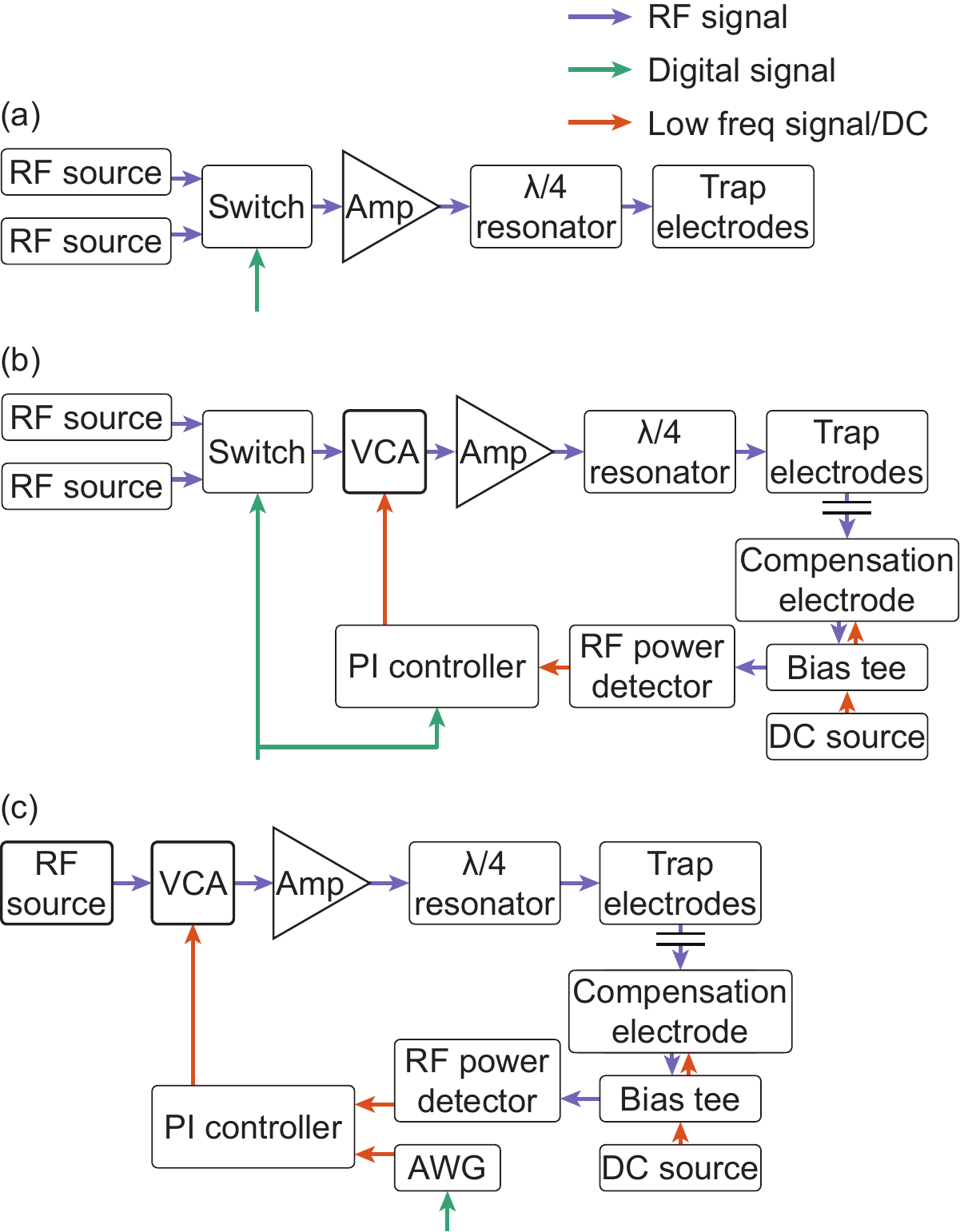}
\caption{Schematics of the different setups we used to control the trap stiffness.
(a)~The trap stiffness is changed by switching between two RF sources which output signals with different amplitudes.
(b)~While the first RF source supplies the signal, the RF voltage on the trap electrodes is actively stabilised by supplying feedback to a voltage-controlled attenuator (VCA).
(c)~The trap stiffness is smoothly changed by using a smoothly changing signal from an arbitrary waveform generator (AWG) as the PI controller's setpoint.
To prevent the figure from being too crowded, we omit the directional couplers placed after the amplifiers, which are used to check that the RF signals are resonant to the resonance circuit.
}
\label{fig_electronics}
\end{figure}

We controlled the trap stiffness during the pulse sequences using three different setups, which are cheap and easy to set up.
They are schematically represented in Fig.~\ref{fig_electronics}.
The simplest setup, in Fig.~\ref{fig_electronics}(a), involved switching between two RF sources\footnote{Analog Devices AD9858} which output signals of different strengths.
On switching the RF source, the RF voltage on the trap electrodes changes to a new value following an exponential decay with time constant $\sim17\,\mathrm{\mu s}$.
This finite time is because the trap electrodes are part of an LC circuit with resonance frequency $2\pi\times 18\,\mathrm{MHz}$ and quality factor $Q\sim300$.
After changing the RF source we typically waited for $50\,\mathrm{\mu s}$ before applying another coherent pulse, to give time for the RF voltage and the trap stiffness to settle.
The switch\footnote{Mini-Circuits ZASWA-2-50DRA+} was controlled by a digital signal, allowing the trap stiffness change to be synchronised with the rest of the experiment.

In ion trap systems which employ a quarter-wave helical resonator \cite{Macalpine1959, Siverns2012} temperature drifts cause the resonance frequency to drift.
This in turn can cause the RF voltage on the trap electrodes and the trap stiffness to drift.
These effects may be exacerbated by the RF power changes during the interferometry sequences, since changing the RF source power changes the power dissipated in the system.
Drifts of the trap stiffness may be mitigated by actively stabilising the RF voltage reaching the trap electrodes \cite{Hempel2014,Johnson2016,Brandl2017}.
Fig.~\ref{fig_electronics}(b) illustrates an extension to the setup in Fig.~\ref{fig_electronics}(a), which allows us to actively stabilise the RF voltage.
Because the micromotion compensation electrodes are capacitively coupled to the trap electrodes, the RF signal leaks onto the compensation electrodes.
We use the RF signal on the compensation electrodes as a proxy for the RF signal on the trap electrodes.
We extract the RF signal from a compensation electrode, measure its amplitude\footnote{Analog Devices AD8361} and feed the amplitude to a PI controller\footnote{Red Pitaya STEMlab 125-14}, which applies feedback to a voltage-controlled attenuator\footnote{Mini-Circuits ZX73-2500-S+}.
Feedback is applied only when the first RF source is used.
When we switch to the second RF source during the micromotion compensation sequence we pause the PI controller's integrator.
This setup was used to obtain most of the results presented here.

The setups in Fig.~\ref{fig_electronics}(a) and (b) involve abrupt changes of the trap stiffness, which causes heating of the ion's motion during the sequences.
Motional heating, in turn, modifies the $|g\rangle \leftrightarrow |e\rangle$ coupling strength and causes pulse area errors.
The setup in Fig.~\ref{fig_electronics}(c) allows for more gradual trap stiffness changes, which lessens the motional heating.
In this setup a signal from an arbitrary waveform generator defines the setpoint of the PI controller\footnote{We use the same Red Pitaya STEMlab 125-14 hardware, but now with the software ``PyRPL'', version ``develop-0.9.3'' \cite{Neuhaus2017, Neuhaus_github}. This software version has the ``differential PID'' option.}.
We smoothly vary the setpoint and in response the trap stiffness varies gradually.
This setup requires only a single RF source.
We typically use the setup in Fig.~\ref{fig_electronics}(b), since it allows for faster sequences, and the short coherence time of our system is a bigger problem than motional heating.

\begin{figure}[ht!]
\centering
\includegraphics[width=\columnwidth]{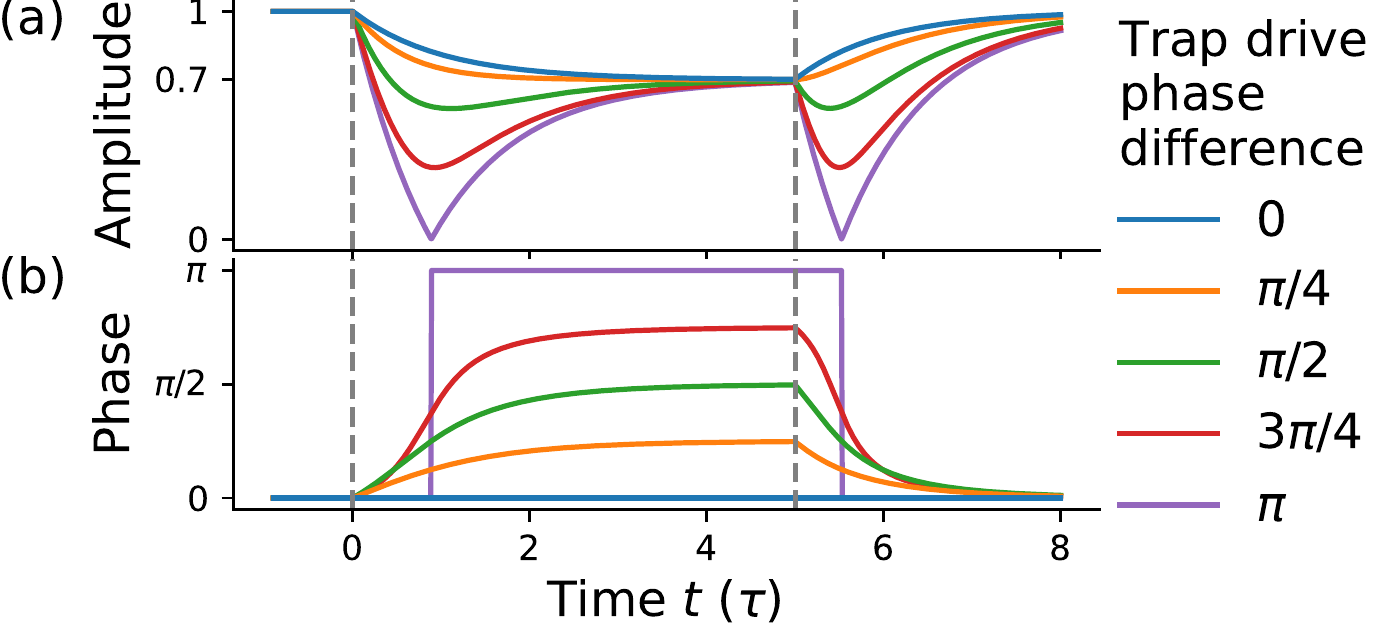}
\caption{Calculated response of the trap's resonator circuit when the RF source is switched at time $t=0$ and reverted at $t=5\tau$, where $\tau$ is the circuit's time constant.
The second RF source supplies a signal with a 30\% lower amplitude.
The response depends on the phase difference between the sources.
When the phase difference is $0$ the response follows an exponential decay.
When the phase difference is high, destructive interference between the incoming RF signal and the RF field inside the resonator causes power to be drawn from the resonator.
}
\label{fig_dds_phase_difference}
\end{figure}

In the two setups involving a switch between two RF sources, its important that the RF signals are in phase, otherwise destructive interference between the incoming RF signal and the RF field inside the resonator will cause power to be drawn from inside the resonator, as shown in Fig.~\ref{fig_dds_phase_difference}.
If too much power is drawn from the resonator circuit, the trap will be momentarily too weak to confine the ion and the ion will be lost.

\section{Binary search algorithm} \label{appendix_rpe_algorithm}
\subsection{Algorithm to combine the results of different measurement sets}
This algorithm is taken from ref.~\cite{Rudinger2017}, we reproduce it here for completeness.
The measurement sets are indexed by $j$.
The $j^\mathrm{th}$ set involves measurements using $M_j=2^{j-1}$, and use Eq.~(\ref{eq_phiT_3}) to return an estimate $\phi_{\mathrm{PD}j}$ of $\phi_\mathrm{PD}$ from within the range $[-\pi/2^{j-1}, \pi/2^{j-1}]$.
The estimates can be combined using the following code, which can be understood with the aid of Fig.~\ref{fig_RPE_explanation}.\\
\texttt{Estimate=0\\
for $j=1$ to $j_\mathrm{max}$:\\
\null\quad L = $\pi/2^{j-1}$\\
\null\quad CurrentEstimate = $\phi_{\mathrm{PD}j}$\\
\null\quad while CurrentEstimate < Estimate - L: \\
\null\quad\quad CurrentEstimate = CurrentEstimate + 2*L \\
\null\quad while CurrentEstimate > Estimate + L: \\
\null\quad\quad CurrentEstimate = CurrentEstimate - 2*L\\
\null\quad Estimate = CurrentEstimate}

\subsection{Remarks about the \textit{Robust Phase Estimation protocol}}
The binary search algorithm we use is based on the \textit{Robust Phase Estimation} protocol introduced in ref.~\cite{Kimmel2015}.
The \textit{Robust Phase Estimation} protocol was designed to be robust in the presence of some types of systematic errors, such as initialization and measurement errors.
In the presence of these errors the phase estimation is \textit{biased}, meaning the accuracy is limited to $\pi/2^{j_\mathrm{max}-1}$, and it does not improve (much) as $N_{j_\mathrm{max}}$ is increased.
Trapped ion systems are not seriously afflicted by initialization or measurement errors, and the phase estimate returned can be \textit{unbiased}, meaning the accuracy improves with $N_{j_\mathrm{max}}$.
This was observed in \cite{Rudinger2017}.

Secondly, the original proposal \cite{Kimmel2015} and the experimental demonstrations (for instance refs.~\cite{Rudinger2017, Meier2019}) of the \textit{Robust Phase Estimation} protocol are mostly concerned with estimation of \textit{rotation angles}; here we estimate a \textit{rotation axis}.

\section{Control phase settings III} \label{appendix_settings_III}
These control phase $\{\theta_j\}$ settings were used in simulations presented in Fig.~\ref{fig_robustness_simulation}
\begin{equation}
\theta_j =\begin{cases}
\tfrac{\pi}{2} &\text{even $j$} \\
-\tfrac{\pi}{2} &\text{odd $j$, } 1<j<M+1 \\
\pi &j=M+1
\end{cases}
\end{equation}
and $\theta_1 \in \{\pi, \tfrac{\pi}{2}\}$, and $M$ is an even integer.
The phase $\phi_\mathrm{T}$ is estimated from
\begin{equation}
\phi_{\mathrm{T}} = \mathrm{arctan2}{\left[ p{\left( \theta_1=\tfrac{\pi}{2} \right)} - \tfrac{1}{2},  p{\left( \theta_1=\pi \right)} - \tfrac{1}{2} \right]  }
\end{equation}

\end{document}